\documentclass[12pt,a4wide]{article}
\usepackage[super,sort&compress,comma]{natbib} 
\usepackage{bm}
\usepackage{times,mathptmx}
\usepackage{amsmath}
\usepackage{amssymb}
\usepackage{graphicx} 
\usepackage[bf,width=11cm]{caption}

\newcommand{\boc}{\mathbf{c}}
\newcommand{\bof}{\mathbf{f}}

\newcommand{\bol}{\mathbf{l}}
\newcommand{\bon}{\mathbf{n}}
\newcommand{\bor}{\mathbf{r}}
\newcommand{\bou}{\mathbf{u}}

\newcommand{\boA}{\mathbf{A}}
\newcommand{\boC}{\mathbf{C}}
\newcommand{\boF}{\mathbf{F}}
\newcommand{\boS}{\mathbf{S}}

\newcommand{\maA}{\mathcal{A}}
\newcommand{\maF}{\mathcal{F}}
\newcommand{\maG}{\mathcal{G}}
\newcommand{\maL}{\mathcal{L}}
\newcommand{\maO}{\mathcal{O}}
\newcommand{\maS}{\mathcal{S}}

\begin{document}
\renewcommand{\thefootnote}{\fnsymbol{footnote}}
\renewcommand\footnoterule{\vspace*{1pt}%
\hrule width 3.4in height 0.4pt \vspace*{5pt}} 
\setcounter{secnumdepth}{5}

\title{Micro- and nanoscale fluid flow on chemical channels$^\dag$}

\author{
Fabian D\"orfler,\textit{$^{a,b}$}
Markus Rauscher,\textit{$^{a,b}$}
Joel Koplik,\textit{$^{c}$} \\
Jens Harting,\textit{$^{d,e}$} and
S. Dietrich$^{\ast}$\textit{$^{a,b}$}}

\maketitle

\abstract{
We study the time evolution and driven motion of thin liquid films
lying on top of chemical patterns on a substrate.  Lattice-Boltzmann
and molecular dynamics methods are used for simulations of the flow
of microscopic and nanoscopic films, respectively.  Minimization of
fluid surface area is used to examine the corresponding equilibrium
free energy landscapes.  The focus is on motion across patterns
containing diverging and converging flow junctions, with an eye
towards applications to lab-on-a-chip devices.  Both open
liquid-vapor systems driven by body forces and confined
liquid-liquid systems driven by boundary motion are considered.  As
in earlier studies of flow on a linear chemical channel, we observe
continuous motion of a connected liquid film across repeated copies
of the pattern, despite the appearance of pearling instabilities of
the interface.  Provided that the strength of the driving force and
the volume of liquid are not too large, the liquid is confined to
the chemical channels and its motion can be directed by small
variations in the geometry of the pattern.
}

\footnotetext{\textit{$^{a}$~Max-Planck-Institut f\"{u}r Intelligente Systeme, Heisenbergstr.\ 3, D-70569 Stuttgart, Germany}}
\footnotetext{\textit{$^{b}$~Institut f{\"u}r Theoretische und Angewandte Physik, Universit\"{a}t Stuttgart, Pfaffenwaldring 57, D-70569 Stuttgart, Germany}}
\footnotetext{\textit{$^{c}$~Benjamin Levich Institute and Department of Physics, City College of the City University of New York, New York, NY 10031, USA, E-mail: koplik@sci.ccny.cuny.edu}}
\footnotetext{\textit{$^{d}$~Department of Applied Physics,
Technische Universiteit Eindhoven, 5612 AZ Eindhoven, The
Netherlands, E-mail: j.d.r.harting@tue.nl}}
\footnotetext{\textit{$^{e}$~Institut f\"ur Computerphysik,
Universit\"at Stuttgart, Pfaffenwaldring 27, D-70569 Stuttgart,
Germany}}
\footnotetext{$^\ast$\textit{E-mail: dietrich@is.mpg.de}}

\section{Introduction}\label{sec:intro}

A major theme in recent technological development has been the
focus on miniaturization and integration.  The prototypical example
is the evolution of microelectronic devices for which speed-up and
cost reduction complement consumer appeal as the main commercial
driving forces for this development.  Analogous considerations have
spawned the field of microfluidics, i.e., the technology of
assembling complex chemical processes into a single miniature
laboratory, as a realization of the so-called lab-on-a-chip concept
\cite{mitchell01,thorsen02,stone04,squires05a,mark10}.  Currently
available devices are made of micron-scale structures but further
miniaturization down to the nano-scale is expected
\cite{abgrall08,vandenberg10a}.

The transition from micro- to nano-scale design and fabrication in
electronics has required new theoretical ideas in the area of
mesoscopic quantum mechanics \cite{imry97,tognetti03,heinzel07}.
Pursuing the analogy, nanofluidic developments require as a new
element the focus on new aspects of fluid mechanics.  On the
nano-scale the macroscopic description of fluids in terms of
classical hydrodynamic equations breaks down.  Boundary layers
dominate, thermal fluctuations and hydrodynamic slip become
relevant, the range of inter-molecular interactions is comparable
to the system size, and the finite size of the constituent
particles has to be taken into account
\cite{eijkel05,mukhopadhyay06,schoch08,kovarki09,bocquet10a}.
Although this presents a major challenge for further
miniaturization there are many ways in which one can make use of
nano-scale specific properties of fluids, e.g., for sorting and
sieving of biomolecules.  Ion channels in biological membranes and
cell walls are another particularly intriguing and inspiring
example
\cite{roth05a,roth06a,jcp_126_2007,roth08,roth09,PRE_82_2010,PRL_105_2010,jcp_135_2011}.

Most commercially available fluidic devices consist of closed
channels formed by a substrate with grooves covered by a plate.
Closed channels offer the advantages of preventing evaporation and
of allowing for pumping by applying a pressure difference between
inlet and outlet.  However, clogging is a serious problem and
cleaning is difficult.  Open microfluidic systems, 
for which
in case of a binary liquid mixture (of, e.g., water and oil)
droplets and rivulets are confined to chemical channels formed by
hydrophilic surface domains on a hydrophobic or less hydrophilic
substrate, have been proposed as a remedy
\cite{gau99,dietrich05,darhuber05,rauscher08a,vanhonschoten10}.
Although one cannot apply a pressure difference between inlet and
outlet, flow in chemical channels can be induced by capillary
forces, i.e., by wicking into channels \cite{darhuber05} or due to
wettability gradients \cite{chaudhury92}, by electrowetting
\cite{quilliet01,zeng04,srinivasan04,mugele05,siribunbandal09},
Marangoni forces due to optically \cite{garnier03,kotz04},
electrically \cite{burns96,farahi04}, or externally generated
temperature gradients, (centrifugal) body forces \cite{koplik06a},
surface acoustic waves \cite{guttenberg05}, or shear flow in a
covering immiscible fluid \cite{rauscher07a}, which would also
prevent evaporation.

A detailed understanding of equilibrium wetting phenomena on
chemically structured substrates is the basis for designing open
microfluidic devices.  In particular the morphology of nonvolatile
fluid droplets on chemical channels has been studied in detail and
morphological transitions have been observed. On 
linear
chemical channels one encounters a transition from bulge-shaped
droplets for large volumes and large equilibrium contact angles on
the channel to elongated rivulets for small volumes and small
equilibrium contact angles on the channel
\cite{gau99,lipowsky00,lipowsky01,brinkmann02,lipowsky05b}.  On
ring-shaped structures there is also a droplet state which covers
the whole ring \cite{lenz01b,porcheron06}.  This richness in
morphological phases is a result of the constant volume constraint
for non-volatile fluids.  Volatile fluids, which are in chemical
equilibrium with their vapor phase, cannot form stable droplets on
homogeneous substrates.  Accordingly, most of the morphologies on
structured substrates, which are stable for non-volatile fluids, 
turn out
to be unstable for volatile fluids: if the pressure in a fluid
configuration decreases with increasing volume the drop will either
grow without limit or evaporate since the vapor acts as a reservoir
with fixed pressure and chemical potential.

However, wetting experiments with volatile fluids are easier to
control than those with non-volatile fluids.  While in the latter
case one must deposit a tiny but well defined amount of fluid on a
substrate, in the first case one can tune the pressure in the
condensed phase by changing the chemical potential of the
reservoir.  In addition, transport through the vapor is much faster
than along the surface, such that equilibration is much faster and
not hindered by pinning of three-phase contact lines.  Equilibrium
wetting phenomena of volatile fluids on homogeneous substrates are
rather well understood \cite{degennes85,dietrich88}: the fluid
forms a homogeneous wetting film which grows in thickness as the
bulk liquid-vapor coexistence line is approached from the gas-side.
The wetting behavior of chemically patterned surfaces has been
studied in detail \cite{bauer99a,bauer99b} and the occurrence of
morphological wetting transitions from very thin to thicker
rivulets have been predicted for 
linear
chemical nano-channels
\cite{bauer99d,bauer00}.
However, only recently it has become possible to manufacture such
structures with negligible height difference between the channel
and the surrounding substrate \cite{checco06,checco09}.

With possible applications as nanofluidic devices in mind, here we
discuss the dynamics of quasi non-volatile droplets (i.e., liquid
droplets with a very small vapor pressure) connected by a
continuous thin 
film, both localized
on a branched chemical channel pattern.  The pattern has the shape
of a ring which is connected periodically to copies of itself. 
This is the simplest geometry which incorporates the key features
of any potential device: one inlet, one outlet, and the splitting
and reconnection of distinct fluid paths.  This pattern has the
further advantage of retaining overall periodicity so that
particles need not be added or removed from the system, which
greatly simplifies all particle-based calculations.  We do not
address dewetting on chemical channels, thus putting aside the
large body of research on dewetting of thin fluid films on
homogeneous and heterogeneous surfaces (see, e.g,
Refs.~\citenum{blossey08,deconinck08,ralston08,quere08,herminghaus08,craster09,mccarty09}
and references therein).

The paper is organized such that in Sect.~\ref{sec_lbe} we review
the lattice-Boltzmann (LB) method with the focus on its application
to simulations of a system with two immiscible fluids.  In
Sect.~\ref{sec_sim} we discuss the shear-cell configuration used in
our calculations, 
in which initially one fluid rests
on top
of a ring-shaped pattern as a liquid ridge, while the second fluid
fills the 
remaining part
of the simulation cell.  The applied LB
method does not take into account thermal fluctuations.  Hence the
results reflect ordinary continuum fluid mechanics and most
reasonably correspond to microscale rather than nanoscale systems.
The simulations generally show that chemical patterning does
successfully direct fluid motion, despite a hydrodynamic instabilty
leading to the appearance of mobile 
non-uniform
bulges or
pearls of the fluid which 
prefers
the chemical channels.  Furthermore, a control
mechanism for influencing fluid motion at a branching 
out
of fluid paths can be inferred from the simulations.

In Sect.~\ref{sec_md} we apply molecular dynamics (MD)
\cite{frenkel02} as the optimal method for describing nanoscale
systems, in order to study the same flow geometry.  These
calculations resemble earlier ones involving a 
linear
(strip) wetting pattern \cite{rauscher07a}.  We again
find that the fluid motion is successfully directed by the pattern.
Remarkably, the results also resemble the LB calculations,
suggesting that chemical patterns provide a very robust means of
flow control.
Although the multicomponent LB calculations presented here cannot
adequately handle a liquid-vapor system, there is no such
limitation for MD \cite{koplik06a}, and we present several examples
of pattern-controlled motion in open systems.  When body force
driven rivulets on sufficiently large patterns are simulated, we
find a limitation for flow control: if too much liquid accumulates
at a region near a junction, surface tension forces may be overcome
by inertia and the liquid may spill off the pattern.

Since surface tension is a dominant force in these systems and
because the fluid behavior at the junctions of the patterns is key
for controlling liquid motion, in Sect.~\ref{sec_freeEn} we study
the interfacial free energy landscape of a fluid droplet near a
junction using the Surface Evolver software package \cite{Brakke}.
Finally, in Sect.~\ref{sec_concl} we present conclusions and an
outlook on future challenges.

\section{The lattice Boltzmann method}\label{sec_lbe}

\begin{figure}
\centerline{\includegraphics[width=11cm]{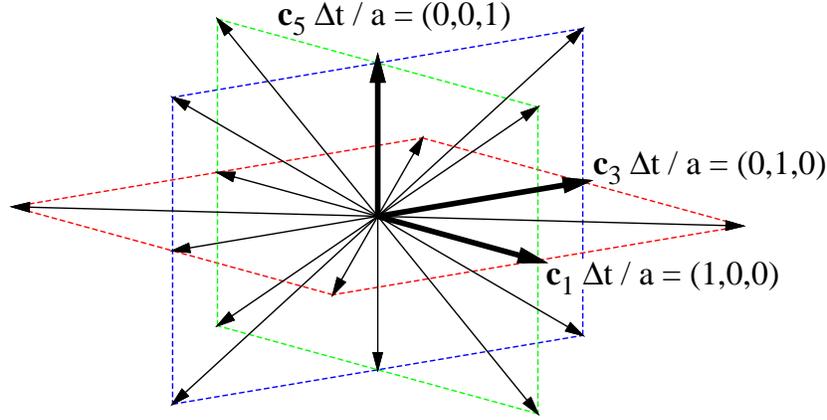}}
\caption{
The structure of the $D3Q19$-lattice.  Starting from the central
site $\bor$ the velocities $\{\boc_1,...,\boc_{18}\}$, allow
particle transport to the neighboring sites located in the planes
(indicated by color) normal to the $\boc_1$, the $\boc_3$, and the
$\boc_5$ direction (dash-dotted vectors).  $\boc_{19}=(0,0,0)$
corresponds to particles resting at $\bor$.  Upon normalization by
means 
of
the lattice constant $a$ and the time step $\Delta t$, the vectors
$\boc_1$, $\boc_3$, and $\boc_5$ form a local, right handed,
orthonormal 
basis so
that
$\bor_\alpha=\bor+\boc_\alpha\Delta t$ renders the lattice sites in
real space.
}
\label{fig_d3q19}
\end{figure}

The LB-method is an indirect solver for the continuum fluid dynamic
equations, which mimics the underlying kinetics of the fluid
particles on a lattice in real space with a discrete set of
velocity degrees of freedom. 
This method is
attractive for the current application since a number of multiphase and
multicomponent models exist which are comparably straightforward to
implement\cite{shan93,shan94,ShanDoolen95,swift95,swift96,gunstensen91,lishchuk03}.
In the following, the LB model used
here is described briefly in order to introduce the parameters
which are crucial for the fluid dynamic description.  A detailed
descussion of the model can be found in the references cited
below.

We use a three-dimensional implementation of the multicomponent
Shan-Chen model (SCM)
\cite{shan93,shan94,nekovee00,harting05,lb_force,HartingJansen2011}
with a 
so-called
$D3Q19$-lattice\cite{qian92}, i.e.,
a three-dimensional cubic lattice in real space with a lattice
constant $a$ and a set of 19 velocity degrees of freedom
$\boc_\alpha$, $\alpha\in\{1,...,19\}$, at any lattice site with a
position vector $\bor$ (see Fig.~\ref{fig_d3q19}).  The state of
the system at a certain time $t$ is given by a set of dimensionless
distributions $f^s_\alpha(\bor,t)$, which upon normalization give
the fraction of particles of species $s$ at a site $\bor$ with the
velocity degree of freedom $\boc_\alpha$, i.e.,
$N^s=\sum_{\{\bor\}}\sum_{\alpha} f_\alpha^s(\bor,t)$ is the total
number of $s$-type particles within the system.  Resting fluid
particles are taken into account by non-zero values of
$f^s_{19}(\bor,t)$ with $\boc_{19}\equiv (0,0,0)$, whereas
$|\boc_{\alpha\neq 19}|\neq 0$.  The particle masses $m^s$ do not
play a physical role in the following, thus we set $m^s\equiv m\
\forall s$, i.e., all particles of all species $s$ have a unit mass
$m$.

Within our implementation of the SCM, the temporal evolution of the
distributions $f^s_\alpha(\bor,t)$ follows the scheme
\begin{multline}
   \frac{f^{s}_\alpha(\bor+\boc_\alpha\Delta t,t+\Delta t) - f^{s}_\alpha(\bor,t)}{\Delta t}
    =\\ -\frac{1}{\lambda^s}
   \left[
      f^{s}_\alpha(\bor,t) - 
         \phi^{s}_{c_T}\left(  |\boc_\alpha-\bar{\bou}^{s}|^2  \right)\big|_{\bor,t}
   \right]
   ,
\label{scm}
\end{multline}
where $\bor$ and $\bor+\boc_{\alpha\neq19}\Delta t$ refer to
neighboring lattice sites, i.e., off-lattice particle positions do
not occur.  The left hand side of Eq.~\eqref{scm} provides the
convection of $f^s_\alpha$ on the lattice along the direction of
$\boc_\alpha$ per 
time unit
$\Delta t$, and the right hand
side effectively models a local relaxation of
$f^s_\alpha$ towards a 
dimensionless
local Maxwellian velocity distribution
$\phi^s_{c_T}$ with a relaxation time scale $\lambda^s$.  The
Maxwellian distribution $\phi^s_{c_T}$ is centered in velocity
space around the velocity $\bar{\bou}^s$ (to be defined below) and
its width 
$c_T=\sqrt{k_B\,T/m}$ is 
proportional to the thermal velocity.
Upon associating the internal energy per particle $\frac{3}{2}\,k_B\,T$
with the kinetic energy $\frac{m}{2}\,(a/\Delta t)^2$ one obtains 
\begin{equation}
   c_T = \sqrt{\frac{k_B\,T}{m}} = \frac{1}{\sqrt{3}}\,\frac{a}{\Delta t} ,
\label{c_T_a}
\end{equation}
which implies $T\propto m\,k_B^{-1}\,(a/\Delta t)^2$, so that
by construction
the SCM is 
isothermal \cite{chen98,Luo00} with the
temperature $T$ fixed by the unit particle mass $m$, the time step
$\Delta t$, and the lattice constant $a$\/.

The zeroth moment of the distributions $f^s_\alpha(\bor,t)$ 
leads 
directly
to the local number density 
of 
species $s$,
\begin{subequations}
\begin{equation}
   \varrho^{s}(\bor,t) = 
	\frac{n^s(\bor,t)}{a^3} =
	\frac{1}{a^3}\,\sum_\alpha f^{s}_\alpha(\bor,t) ,
\label{loc_dens}
\end{equation}
where $n^s(\bor,t)=\sum_\alpha f^{s}_\alpha(\bor,t)$ renders the
number of particles 
of species $s$
at the site $\bor$.
The local 
velocity field
$\bou^{s}(\bor,t)$ for species $s$
is defined by the first velocity moment and the so-called Shan-Chen
acceleration $\boF^s$,
\begin{equation}
   \bou^{s}(\bor,t)
   = \frac{1}{n^s}
   \left(
      \textstyle\sum_\alpha\boc_\alpha\,f^{s}_\alpha + \frac{1}{2}\,\boF^s\Delta t
   \right) .
\label{loc_vel}
\end{equation}
\end{subequations}
The last term is needed in order to compensate for discretization
artefacts and to obtain Navier-Stokes equations on the fluid
dynamic level \cite{lb_force}.  The corresponding total quantities
are given by a summation over all types:
$\varrho(\bor,t)=\sum_s\varrho^s$ and
$\bou(\bor,t)=\varrho^{-1}\sum_s\varrho^s\,\bou^s$\/.

The key feature of the SCM is the acceleration \cite{shan93}
\begin{equation}
   \boF^s(\bor,t) =
   -\psi^s\big|_{(\bor,t)}
      \sum_{\{\tilde{s}\}}\,\maG^{s\tilde{s}} 
         \,\sum_{\tilde{\alpha}}\psi^{\tilde{s}}\big|_{(\bor+\boc_{\tilde{\alpha}}\Delta t,t)}\,\boc_{\tilde{\alpha}}\Delta t
   ,
\label{scm_force}
\end{equation}
which is often interpreted as the local effective acceleration of a
particle of type $s$ at the lattice site $\bor$ due to the presence
of particles at the neighboring sites
$\bor+\boc_{\tilde{\alpha}}\Delta t$\/.
The coupling strength $\maG^{s\tilde{s}}$ defines attraction
(negative sign of $\maG^{s\tilde{s}}$) or repulsion (positive sign
of $\maG^{s\tilde{s}}$) between two particles of type $s$ and of
type $\tilde{s}$.
The so-called pseudo density
$\psi^s|_{(\bor,t)}=1-\exp[-n^s(\bor,t)]$ is determined by the
local partial particle number $n^s(\bor,t)$\/.
The acceleration $\boF^s$ enters the evolution scheme given by Eq.~\eqref{scm} via the velocity $\bar{\bou}^s$, which determines
the center of the Maxwellian distribution $\phi^s_{c_T}$ in velocity space,
\begin{multline}
  \bar{\bou}^s(\bor,t)
   = \bou^s\ +\ \frac{\boF^s}{n^s}\lambda^s
   \\= \frac{1}{n^s}
  \left[
     \textstyle\sum_\alpha\boc_\alpha f^s_\alpha\ +\ \left( \tfrac{1}{2}\Delta
	  t + \lambda^s \right)\,\boF^s
  \right].
\label{loc_vel_eq}
\end{multline}

The expressions for $\bou^s$, $\boF^s$, and $\bar{\bou}^s$ given by
Eqs.~\eqref{loc_vel},~\eqref{scm_force}, and~\eqref{loc_vel_eq},
respectively, are constructed such that in the large scale limit
one obtains fluid dynamic equations with a non-ideal equation of
state. 
(In the case of a one-component system for a negative sign of $\maG^{ss}$, i.e.,
attraction among particles of the same species, the pseudo density
$\psi^s$ as given above leads to a van-der-Waals loop.)
One can
show that this approach is equivalent to treating the acceleration
$\boF^s$ formally as an external field.

In the simulations, we minimalisticly mimic the kinetics of a
system consisting of an $o$-type fluid (oil) covering a $w$-type
fluid (water) rivulet so that each fluid is in contact with certain
parts of an $r$-type substrate (rock, or more generally, solid).
The model is minimalistic in the sense that oil and water particles
are distinguished by a repulsive interaction only, i.e.,
$\maG^{ow}>0$ with equal relaxation time scales for both oil and
water, $\lambda^o=\lambda^w\equiv\lambda$.
The rock particles are pinned to their lattice sites, i.e.,
$f^r_{19}(\bor,t)\equiv$ const and
$f^r_{\alpha\neq19}(\bor,t)\equiv0$ for all times $t$,
respectively.
Together with bounce back boundary conditions this effectively
results in a hydrophobic solid boundary 
if $f_{19}^r >0$ and $\maG^{rw}>0$\/.

In order to induce oil-water phase separation and thus to facilitate wetting, 
our
choice for the coupling constants
$\maG^{s\bar{s}}$ is 
\begin{equation}
   \maG^{oo} = \maG^{ww} = \maG^{or} \equiv 0, \quad 
   \maG^{ow} = \maG^{rw} \equiv  \maG > 0 .
\label{scm_setup}
\end{equation}
The phase separation of oil and water 
is
driven by the parameter $\maG$, and the wetting behavior for a
given $\maG$ depends on the local, partial numbers densities of
oil, water, and rock particles, $\varrho^o$, $\varrho^w$, and
$\varrho^r$, respectively. 
With this choice of parameters,
phase separation is almost 
complete
and the concentration of oil in
the water phase and vice versa is negligible (smaller than $0.2$\%\
in weight).

\subsection{The fluid dynamic level}

In the strong segregation limit, within the water or the oil phase the
fluid dynamic equations for the 
majority species
$s\in\{o,w\}$,
i.e., for the quantities $\varrho^s$ and
$\bou^s$ defined by Eqs.~\eqref{loc_dens} and \eqref{loc_vel},
respectively, can be obtained via a systematic Chapman-Enskog 
analysis of Eq.~\eqref{scm}
\cite{chen92,ShanDoolen95,chen98,lb_force}.
With $\partial_t=\partial/\partial t$, $\partial_i=(\bm{\nabla})_i$ and
$u^s_i=(\bou^s)_i$ for $i=1,2,3$, and upon neglecting terms of
$\maO\left((\bou^2/c_T)^{k>2}\right)$, i.e., within the regime of
sufficiently 
small
Mach numbers $(\bou^s/c_T)^2$, the continuity equation is
\begin{subequations}
\begin{equation}
   \partial_t\varrho^s + \partial_i(\varrho^su^s_i)\ =\ 0
   \quad ,
\label{scm_ncons}
\end{equation}
and the balance of momentum reads
\begin{equation}
   m\varrho^s\left(\partial_t + u^s_j\partial_j\right) u^s_i
    \approx 
   -\partial_jp^s\delta_{ij}
    + \partial_j\,\eta^s\left(\partial_ju^s_i + \partial_iu^s_j\right) .
\label{scm_flow}
\end{equation}
\label{scm_fluid_dyn}
\end{subequations}
In the limit of nearly complete phase separation, the density
of the minority phase is negligible and therefore it does not
contribute to the momentum balance. 
These are the compressible Navier-Stokes equations, which imply
number conservation and conservation of the total momentum 
as the sum of all momenta of all particles (Upon
$(\bou^s/c_T)^2\rightarrow 0$ the incompressible regime is
approached asymptotically because the Chapman-Enskog analysis renders
both $\bm{\nabla}\varrho$ and $\bm{\nabla}\cdot\bou^s$ to be of
$\maO\left((\bou^s/c_T)^2\right)$.)

For a single component and single phase system the local pressure
can be calculated using the equation of state
$p=m\,c_T^2 \, \rho = k_B\,T\,\rho$ of an ideal gas. However, for a
binary mixture and for the choice of
parameters given in Eq.~\eqref{scm_setup} non-ideal contributions
to the pressure come into play and one obtains
\begin{equation}
\frac{p}{m\,c_T^2} = \rho^o + \rho^w + \frac{(\Delta
t)^2}{a^3}\sum_{s,\bar{s}\in\{o,w\}}
\maG^{s\bar{s}}\psi^s\psi^{\bar{s}}.
\label{scm_eos}
\end{equation}
With the choices given by Eq.~\eqref{scm_setup} and a coupling
parameter $\maG$ strong enough to induce nearly complete oil-water
phase separation, the partial pressure of the minority species is
negligible. The partial pressure of the majority species $s$
in the homogeneous $s$-phase resembles that of an ideal gas,
\begin{equation}
   p^s  =  \varrho^s\,k_B\,T ,
\label{scm_eos_id}
\end{equation}
because a vanishing pseudo density $\psi^{\bar{s}}$ of the minority
species $\bar{s}$ renders the non-ideal part in Eq.~\eqref{scm_eos}
negligible.  Non-ideal behavior appears only within the interfacial
region separating the two distinct homogeneous phases (with
majority species $w$ and $o$, respectively), where the
Chapman-Enskog analysis is not applicable.  This implies that
interfacial tensions and contact angles have to be determined from
simulation results.  At present there are no explicit formulae
which allow one to calculate them directly from the simulation
parameters.

In a nearly completely phase separated system
the shear viscosity in the homogeneous $s$-phase reads
\begin{equation}
  \eta\approx  \eta^s =
  m\,\varrho^s\,c_T^2\,\left(\tau-\tfrac{1}{2}\right)\,\Delta t
\label{scm_visc}
\end{equation}
with the dimensionless relaxation time $\tau=\lambda/\Delta t$.
We note that, first, $\eta^s$ depends linearly on the local number
density $\varrho^s$, which is a property of dilute fluids,
second, that $\eta^s$ changes upon varying the timestep $\Delta t$,
and third, that $\eta^s$ does not depend on the coupling parameters
$\maG^{s\bar{s}}$ (because in Eq.~\eqref{scm_setup} the self-coupling is set to zero)\/.

\section{Lattice Boltzmann simulations}\label{sec_sim}

\begin{figure} 
\centerline{\includegraphics[width=9cm]{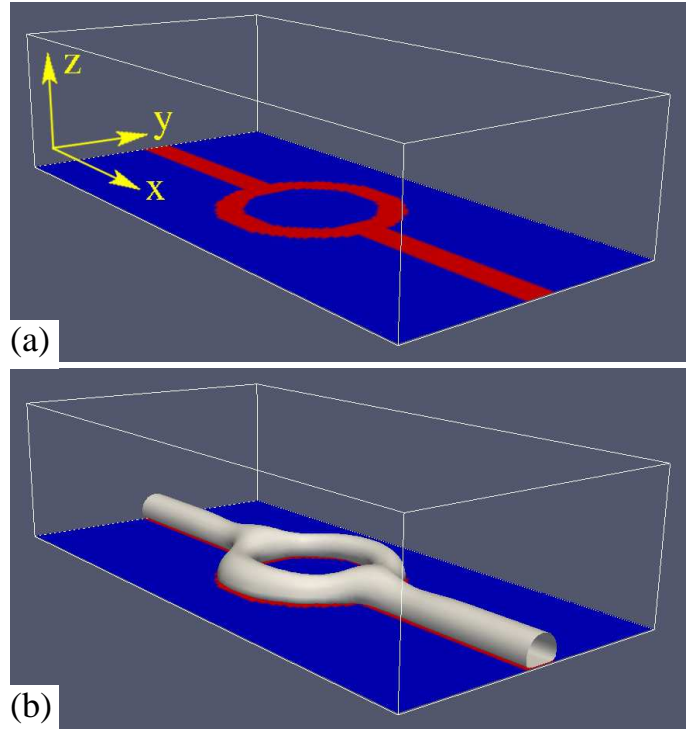}}
\caption{
(a)
We have used a simulation box with dimensions
$128a\times64a\times32a$ 
in $x$-, $y$-, and $z$-direction
and with the substrate in the $xy$-plane
as the bottom wall.  The chemical pattern provides hydrophilic
areas, marked in red, embedded in a hydrophobic substrate, marked
in blue.  The clipping of hydrophilic and hydrophobic domains is
sharp, i.e., the transition between the red and the blue domains
occurs within one lattice spacing $a$.  In the symmetric case all
the hydrophilic sections have a uniform width $w=6a$, and the
outer radius of the ring is $20a$.  In the asymmetric case the
outer radius of the front branch (the branch with smaller values
of $y$) is narrowed by one $a$.  The pattern junctions are termed
the upstream and the downstream junction with respect to the
direction of the shear velocity $\bou^{o,w}\big|_\text{top
layer}=(v_\text{shear}\ge 0,0,0)$ in $x$-direction.  (b) Morphology 
of a water rivulet on a symmetric pattern at early stages (100
$\Delta t$) for $\theta_\text{phil}\approx 80^\circ$, 
$\theta_\text{phob}\approx 140^\circ$, and
$v_\text{shear}/c_T\approx 0.12$, displayed as the isosurface $\maA$ of
half the density in bulk water.  For a description of the initial
configuration ($t=0$) see the main text.  The onset of the
pearling instability is already visible in the form of small
bulges above the pattern junctions.  }
\label{fig_geo_lbe}
\end{figure}

\begin{figure} 
\centerline{\includegraphics[width=9cm]{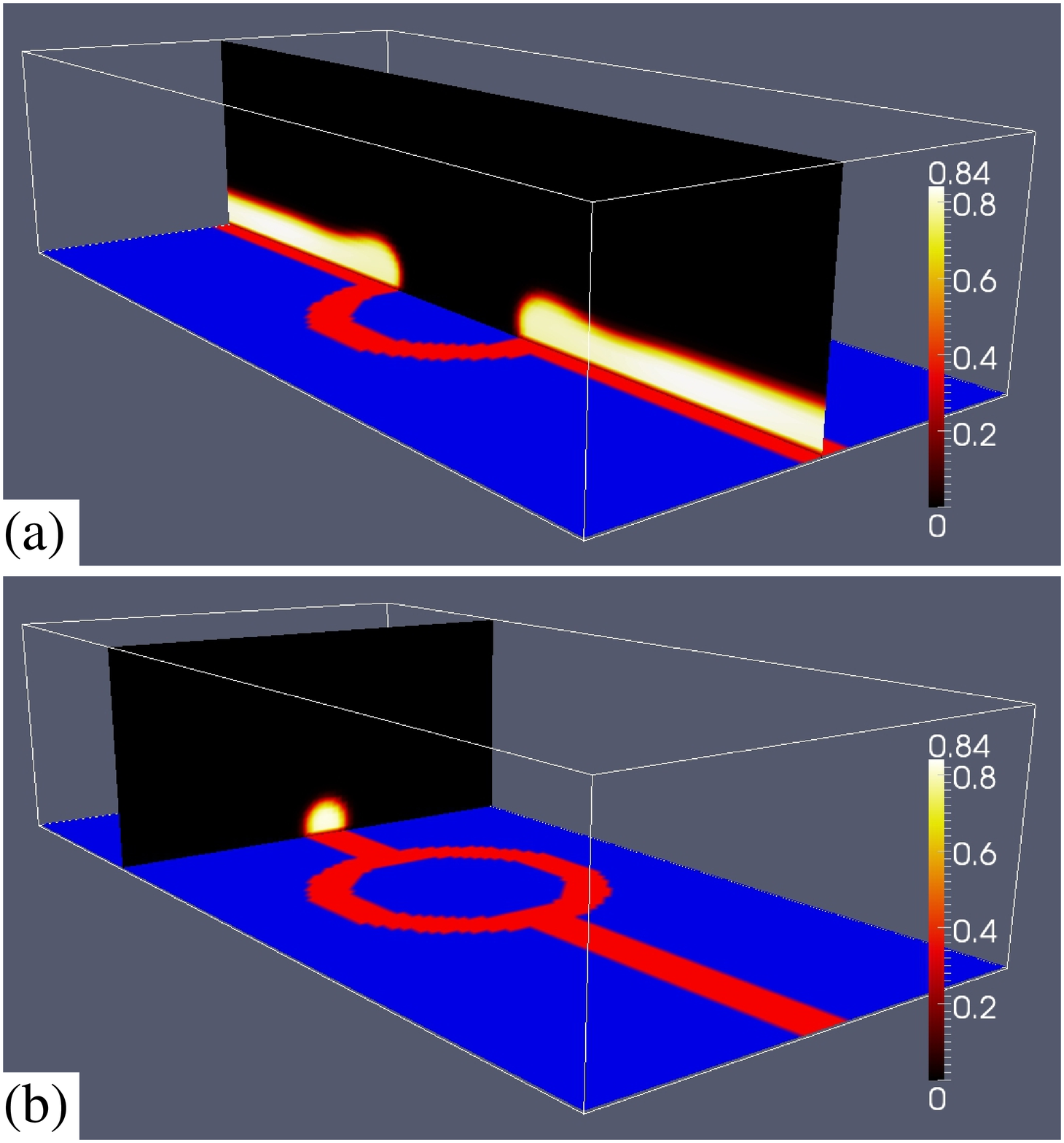}}
\caption{
Cuts of the density profile of the configuration shown in
Fig.~\ref{fig_geo_lbe}(b), along
(a) and perpendicular (b) to the $xz$-symmetry plane.  The color code refers to the
water density $\varrho^w$ 
in units of $a^{-3}$
with a bulk
value (i.e., in the
interior of the rivulet) of $\varrho^w\approx 0.84\,a^{-3}$.  The
water density drops from the bulk value to almost zero
within
a
range of about $4a$.  The same holds for the oil 
density; however,
the bulk value of the oil density (i.e., far from the
substrate) is
$\varrho^o\approx0.74\,a^{-3}$,
$\theta_\text{phil}\approx
80^\circ$ and $\theta_\text{phob}\approx 140^\circ$.
}
\label{fig_cuts_lbe}
\end{figure}

The chemical pattern is implemented via a lateral variation of the
fixed rock density $\varrho^r$ leading to channel-like hydrophilic
surface domains with a certain width $w$ embedded in a hydrophobic
substrate with contact angles $\theta_\text{phil}<90^\circ$ and
$\theta_\text{phob}\approx 140^\circ$, respectively.  (The contact
angle is defined as the angle of the oil-water interface at a
homogeneous rock surface for a macroscopicly large water droplet.)
We have investigated two types of patterns: a symmetric pattern
with $xz$-mirror symmetry, and an asymmetric pattern with broken
$xz$-mirror symmetry (see Fig.~\ref{fig_geo_lbe}(a)).  The
simulation box is designed as a shear cell, which means that a
no-slip boundary condition is imposed at the substrate, 
both
lateral boundary conditions are periodic, and a constant flow
velocity aligned parallel 
to
the $x$-direction is imposed within the top layer of
lattice sites in the simulation box, i.e.,
$\bou^{o,w}(\bor,t)\big|_\text{top\,layer}\equiv(v_\text{shear},0,0)$.

For both the symmetric and the asymmetric pattern, we have run
simulations for two different hydrophilic contact angles,
$\theta_\text{phil}\approx30^\circ$ and
$\theta_\text{phil}\approx80^\circ$, and for two different shear
velocities, $v_\text{shear}/c_T=0.12$ and
$v_\text{shear}/c_T=0.17$.  The initial fluid geometry is given by
a water rivulet covering the pattern with a square cross section
$w\times w$ and the rest of the simulation box filled with oil,
i.e., for the box dimensions given in
Fig.~\ref{fig_geo_lbe} one has $V/w^3\approx32$ for
the water volume $V$.  Both water and oil have
homogeneous initial number densities,
$\varrho^{o,w}=\varrho=0.7\,a^{-3}$, and the spatial variation of
the densities between the water and the oil phases occurs within
one lattice spacing $a$.  The coupling strength within the
acceleration $\boF^s$ (Eq.~\eqref{scm_force}) is
$\maG=0.20/\lambda^2$ and the dimensionless relaxation time is
$\tau=\lambda/\Delta t=1.0$, which leads to an oil-water surface
tension\cite{Schmie2011}  $\gamma\approx 0.07\,m/\lambda^2$ and a
shear viscosity $\eta^{o,w}=m\varrho c_T^2\lambda/2$ for both oil
and water (see Eqs.~\eqref{c_T_a} and~\eqref{scm_visc} for
$\tau=1$).

With $\tau =1$ the
Reynolds number and the capillary number
for the water rivulet are 
$\text{Re}
=m\varrho^wwv_\text{shear}/\eta^w=\sqrt{12}\,(w/a)\, (v_\text{shear}/c_T)$ and 
$\text{Ca}=\eta^w\,v_\text{shear}/\gamma=[(c_T\,\lambda)^3\varrho^w/(2\,\gamma\,\lambda^2)]\,
(v_\text{shear}/c_T)
=
\left[m\,\varrho^w\,a^3/(\sqrt{108}\,\gamma\,\lambda^2)\right]\,(v_\text{shear}/c_T)
$,
i.e., due to $\tau=1.0$ and
$w/a=6$,
\begin{equation}
   \text{Re}
   \approx 20.8\,\frac{v_\text{shear}}{c_T},\quad 
   \text{Ca}
   \approx  0.96\,\frac{v_\text{shear}}{c_T} .
\end{equation}
According to our choices given above, the
ratio
$v_\text{shear}/c_T$ is of $\maO(10^{-1})$, 
so that
Re is of $\maO(10^0)$ and Ca is of $\maO(10^{-1})$. 
This
means that we
have laminar flow with a strong 
influence of the
surface tension on the evolution of the oil-water interface.
The 
temporal
evolution of the system is followed by monitoring 
the
isosurface of the water density corresponding to half its bulk
value.

\subsection{General features}\label{ssec_lbe_general}

The initially square-shaped water ridges evolve very fast (within
less than 100 $\Delta t$) towards a circular cross-sectional shape.
Oil and water are separated into coexisting thermodynamic
bulk phases, i.e., there is an oil bulk phase with a very low
density of water particles (smaller than
$0.2$\%\ in weight) and vice versa.  The oil-water interface has a well-defined
width of about $4a$ (see Figs.~\ref{fig_geo_lbe}$\,(b)$
and~\ref{fig_cuts_lbe}).  The water and oil bulk densities are
somewhat above the initial values: 
$\varrho^w\approx 0.84\,a^{-3}$ and
$\varrho^o\approx 0.74\,a^{-3}$ compared to the initial
values $\varrho^w=\varrho^o=0.7\,a^{-3}$. This
is a consequence of the Laplace
pressure (recall Eq.~\eqref{scm_eos_id}) in combination with the finite 
volumes of the emerging interfacial regions.

The water ridges are unstable with respect to pearling
\cite{koplik06a}, whereupon the dynamics of the pearling process
for a given non-zero shear velocity $v_\text{shear}$ is controlled
by the values of Ca and 
$\theta_\text{phil}$.
is driven by the oil-water surface tension and is hindered by
viscosity and the attractive fluid-substrate interactions, which
means that the pearls preferentially form at the pattern junctions
and the pearling is slower and less pronounced for the case
$\theta_\text{phil}\approx30^\circ$ compared to the case
$\theta_\text{phil}\approx80^\circ$.  Due to compressibility
(recall Eq.~\eqref{scm_fluid_dyn}), the water
density in the interior
decreases with a decreasing mean curvature of
the oil-water interface, i.e., $\varrho^w<0.8\,a^{-3}$ 
inside
big droplets enclosing the vast majority of water in the system and
$\varrho^w>0.9\,a^{-3}$ 
inside
very small droplets.

\subsection{$\theta_\text{phil}\approx80^\circ$}

\begin{figure*} 
\includegraphics[width=\linewidth]{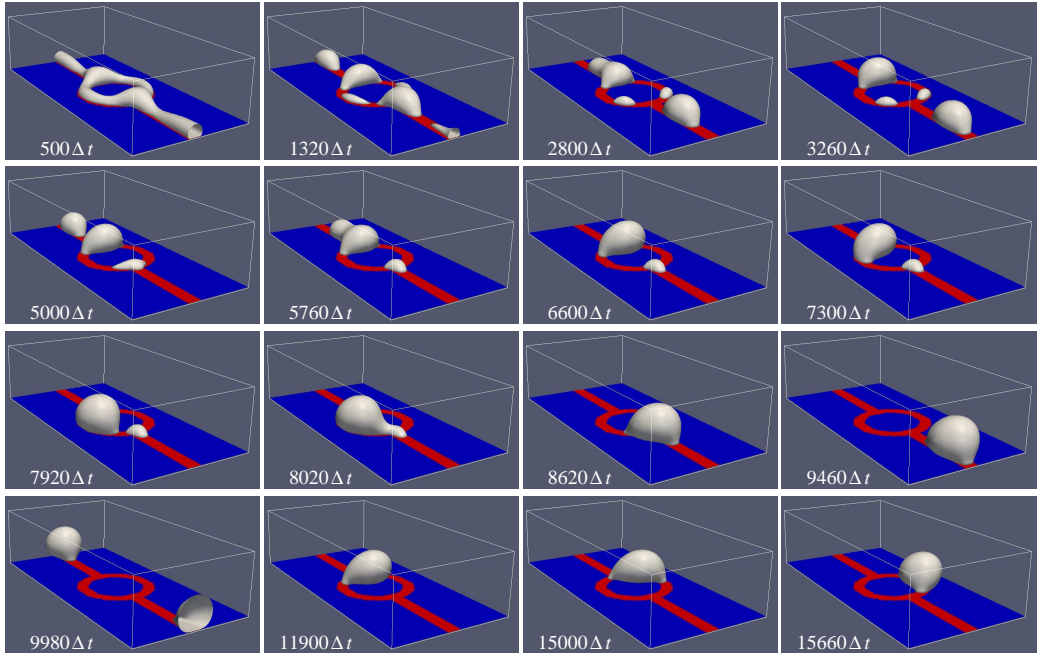} 
\caption{
Simulation snapshots for $\theta_\text{phil}\approx
80^\circ$,
$\theta_\text{phob}\approx140^\circ$,
and
$v_\text{shear}/c_T=0.12$ on a symmetric pattern.  The pictures
are arranged in 
consecutive
rows with the flow of time
from left to right.  The pearling at early times, the aggregation
and merger of droplets at the junctions, as well as the random
droplet behavior at the upstream junction (i.e., whether turning
right or left) can be seen.  The randomness is introduced by
numerical round-off errors in the 
simulation
code.  Note
the periodic boundary conditions in 
$x$-direction. 
}
\label{fig_a80_v02}
\end{figure*}

\begin{figure*}
\includegraphics[width=\linewidth]{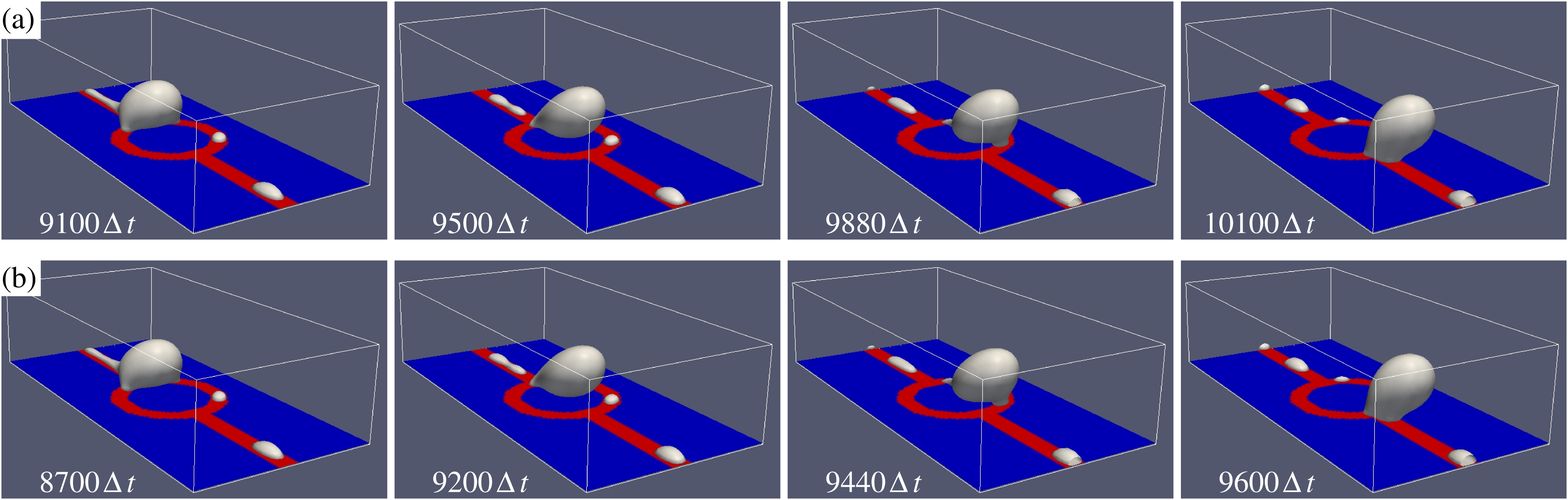}
\caption{
Simulation snapshots for $\theta_\text{phil}\approx
80^\circ$,
$\theta_\text{phob}\approx140^\circ$,
and
$v_\text{shear}/c_T=0.17$ on a symmetric (a) and an asymmetric
pattern (b)\/.  In the symmetric case the droplet is driven off
the pattern, whereas in the asymmetric case it is still guided by
the wider rear branch of the pattern. 
}
\label{fig_a80_v03}
\end{figure*}

Figure~\ref{fig_a80_v02} shows the time evolution for
$v_\text{shear}/c_T=0.12$ and $\theta_\text{phil}\approx80^\circ$.
At early times, the water transport due to pearling dominates over
the transport due to shear flow.  This means that whole water
droplets move along the wetting pattern in tandem with their
formation by the pearling process.  The figure appears to show
isolated droplets separated by a waterfree substrate, but this is
an artifact of the plotting procedure.  In fact, a layer of water
with about one to two $a$ in thickness and with approximately $1\%$
of the water bulk density is permanently adsorbed at the pattern.
Density fluctuation due to numerical round-off errors influence the
fluid behavior at the symmetric upstream junction, i.e., the
droplets randomly choose one of the two branches to move
downstream.  At an asymmetric upstream junction, due to surface
tension forces the wider rear branch is preferred.  Bigger droplets
sample more of the linear shear flow profile and therefore are
driven faster than smaller ones.

Figure~\ref{fig_a80_v03} shows the fluid behavior for the higher
shear rate $v_\text{shear}/c_T=0.17$.  In the symmetric case, at
this rate big droplets are finally driven off the pattern at the
upstream junction, whereas the big final droplet is still guided by
the pattern in the asymmetric case, yet with a significant spillage
onto the hydrophobic domain ($\theta_\text{phob}\approx
140^\circ$).  Due to the enhanced shear drive, tails grow out of
the moving droplet, which eventually break up into pearls.  (This
phenomenon might be related to the formation of Landau-Levich films
in coating problems.)


\subsection{$\theta_\text{phil}\approx30^\circ$}

\begin{figure*}
\includegraphics[width=\linewidth]{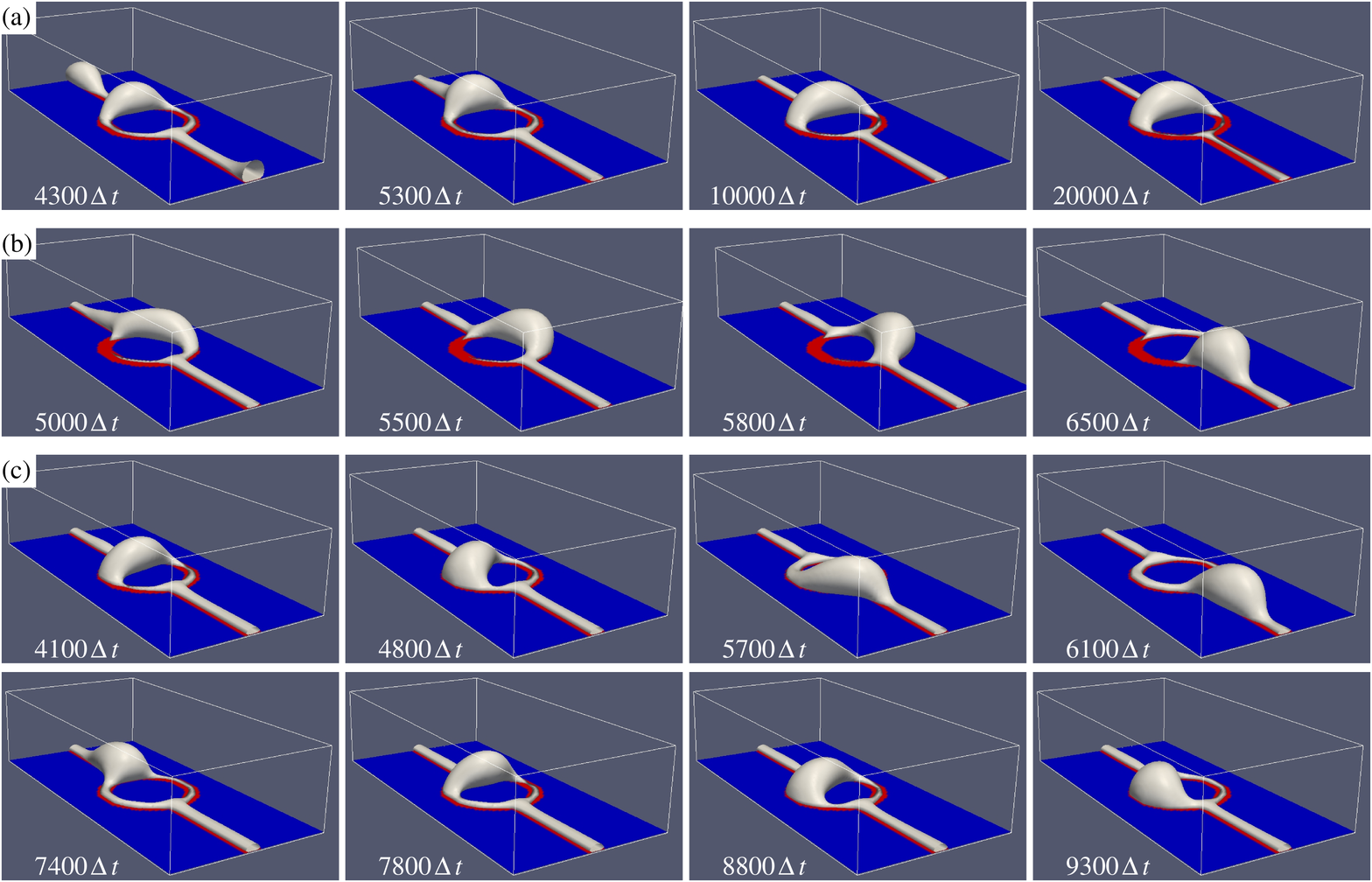}
\caption{
Simulation snapshots for $\theta_\text{phil}\approx
30^\circ$,
$\theta_\text{phob}\approx140^\circ$,
and
$v_\text{shear}/c_T=0.12$ on a symmetric (a) and an asymmetric
pattern (b)\/.  While the droplet gets stuck permanently at the
symmetric upstream junction there is no hang up in the asymmetric
case.  Due to surface tension, the droplet favors the wider rear
branch.  (c) Simulation snapshots for
$\theta_\text{phil}\approx 30^\circ$ and
$v_\text{shear}/c_T=0.17$ on a symmetric pattern.  The droplet at
the upstream junction randomly chooses one of the branches.  Due
to the periodic boundary 
conditions it always takes, however, the
same path in any consecutive passage because the
mirror symmetry of the system remains broken by an unequal
coating of the pattern by water.  The droplet leaves behind a
significant amount of water on the channel.
\label{fig_a30} }
\end{figure*}

Figures~\ref{fig_a30}(a) and (b) show the results for the shear
rate $v_\text{shear}/c_T=0.12$.  On the symmetric pattern, a single
droplet gets stuck permanently at the upstream junction, whereas
there is no permanent hang up on the asymmetric pattern.  The
enhanced water-rock interaction leads to a water coating of the
channel sections the droplet has run over, i.e., a water coverage
far beyond the minimal adsorption mentioned above.

For the higher shear rate $v_\text{shear}/c_T=0.17$
Fig.~\ref{fig_a30}(c) shows that no permanent hang up occurs at a
symmetric upstream junction.  Instead, numerical round-off errors
influence the fluid behavior as already described for
$\theta_\text{phil}\approx 80^\circ$ and $v_\text{shear}/c_T=0.12$.
However, after the first passage the $xz$-mirror symmetry of the
system remains broken by an unequal water coating, and hence the
droplet path is predetermined for any consecutive passage by the
film left behind the droplet.  In the asymmetric case, the
situation is qualitatively the same as the one above for
$v_\text{shear}/c_T=0.12$.


\subsection{Flow control options}

The upstream junction where the inlet channel branches out in flow
direction turns out to be the crucial component of the pattern,
because -- provided that the surface tension forces are strong
enough to keep the droplet on the pattern -- the droplet behavior
is determined by symmetry.  Droplets get stuck at a symmetric
junction since the mirror symmetry of the pattern naturally
translates into a mirror symmetry of the droplet morphology, and a
splitting of the droplet into two smaller ones is prevented by
surface tension.  In contrast, surface tension forces drive the
droplets onto the wider rear branch of an asymmetric junction.

This might be exploited in order to direct fluid motion and control
droplet throughput on potential microfluidic devices: A droplet
typically running on the wider rear branch of the asymmetric
pattern could be occasionally forced to take the front branch by
means of a time-dependent, local variation of the wettability near
the junction generated, e.g., by electric pulses or heating.

\section{Molecular dynamics simulations}\label{sec_md}

\subsection{Methods}\label{sec_md_methods}

The MD simulations involve generic viscous fluids
and a crystalline solid substrate constructed of atoms interacting
via Lennard-Jones potentials.  The shape of the wetting pattern on
the solid is similar to the 
pattern used above for the LB simulations
and the
computational method is identical to that used in two previous
papers.  The case of 
immiscible binary liquid mixtures
in a shear cell follows Ref.~\citenum{rauscher07a}, while related
calculations involving an open geometry with a liquid ridge
of a one-component fluids
in contact with its vapor is based
on Ref.~\citenum{koplik06a}.  The pair 
potentials are of the
Lennard-Jones form,
\begin{equation}
  \Phi_{\rm LJ}(r) =
  4\,\epsilon\, \left[ \left( \frac{r}{\sigma} \right)^{-12} -
  c_{s\tilde{s}}\, \left( \frac{r}{\sigma} \right)^{-6}\ \right] ,
\end{equation}
with the atomic core size $\sigma$ and the potential depth
$\epsilon$ as the characteristic scales of length and energy,
respectively.  The fluid atoms of all species have the common mass
$m$, so that $t_0=\sigma\,\sqrt{m/\epsilon}$ gives a characteristic
time scale.
The interaction is cut off at $r_c=2.5\,\sigma$ and shifted by a
linear term so that the force vanishes smoothly there.  The
coefficient $c_{s\tilde{s}}$ is used to vary the strength of the
attractive interaction between atomic species $s$ and $\tilde{s}$.

In the shear cell case the interaction strength coefficients have
the standard value 1.0, except that $c_{s\tilde{s}}=0$ if $s$ and
$\tilde{s}$ refer to 
atoms of
different liquids, or
of
the inner liquid (the ``water'') and 
of
the
regions outside the
pattern, or 
of
the outer liquid (the
"oil") and of the regions 
of
the pattern.
With this, the two liquids are 
de facto
immiscible and the inner liquid wets the pattern completely
($\theta_\text{phil}=0^\circ$), while the regions 
outside
the pattern are 
completely
nonwetting ($\theta_\text{phob}=180^\circ$).

In the open case \
with a one-component fluid,
the
liquid-liquid and 
the
solid-solid interaction strength
coefficients are $1.0$, and the liquid-solid coefficients are $1.0$
for solid atoms within the pattern (leading to a contact angle
$\theta_\text{phil}=0^\circ$) and either $0$
(``nonwetting'', $\theta_\text{phob}=180^\circ$) or $0.75$
(``partial wetting'', $\theta_\text{phob}\approx90^\circ$) for the
solid atoms outside the pattern.
In this open case the
liquid consists actually of linear chains of four atoms
each, joined by nonlinear (FENE) springs, so as to reduce the vapor
pressure and to provide a reasonably sharp interface. 

The 
atoms of the solid substrate
are not free, but are tethered to lattice sites by a stiff linear
spring with the force law $\bof=-k\,(\bor-\bor_0)$, where
$k=100\,\epsilon/\sigma^2$ and $\bor$ and $\bor_0$ are the position
of a wall atom and 
of
its lattice position, respectively.
The wall atoms have a mass $m_w=100\,m$ so that the characteristic
frequency 
$\sqrt{k/m_w}=1/\tilde{t}_0$
of the wall atom
motion is less than that of dimer harmonic oscillations about the
LJ potential minimum, which is approximately 
$7.6/\tilde{t}_0$.

In the shear cell case, the simulation involves 120,040
``water''
atoms, 1,799,960 
``oil''
atoms and 154,880 
substrate
atoms, in a box of dimensions $273.6\,\sigma\times171.0\,\sigma$ in
the substrate plane, with a height of $54.72\,\sigma$.  The 
of the wetting strip is $17.1\,\sigma$ and the outer radius of
the circular arc is $68.4\,\sigma$.

\begin{figure}
\centerline{\includegraphics[width=11cm]{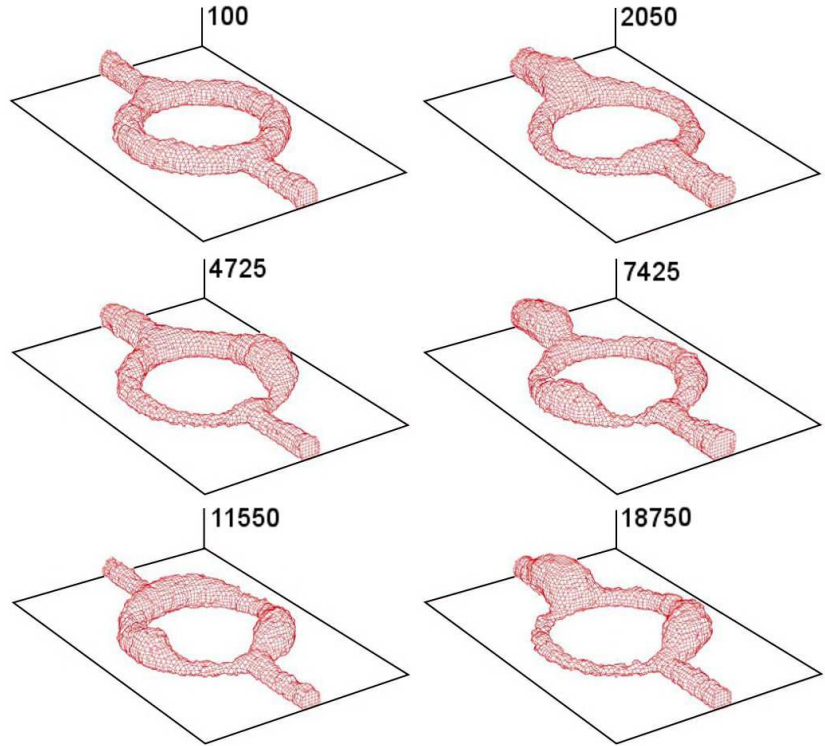}}
\caption{ \label{fig:md_open} 
Sequence showing the evolution of a uniform,
one-component
liquid ridge on top of a symmetrical wetting pattern in contact
with its vapor.  The surface represents the liquid-vapor
interface $\maA$, and motion is driven by a body force of magnitude
$0.001\, m\,\sigma/t_0^2$.  The interaction coefficient
$c_{liquid,rock}$ is 1.0 inside the pattern and 0 outside;
$\theta_\text{phil}=0^\circ$ and $\theta_\text{phob}=180^\circ$\/.
The other simulation parameters are given in
Sect.~\ref{sec_md_methods}.
}
\end{figure}

\begin{figure}
\centerline{\includegraphics[width=11cm]{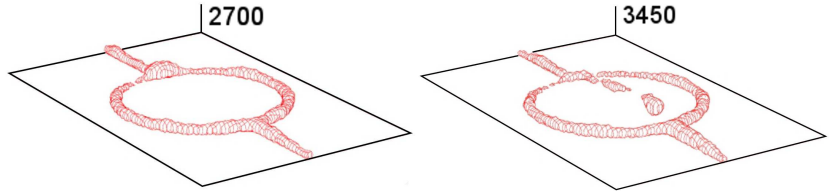}}
\caption{ \label{fig:md_big_nw}
Behavior of a symmetrical junction pattern with twice the
dimensions of Fig.~\ref{fig:md_open}.  The other simulation
parameters are the same as in Fig.~\ref{fig:md_open};
$\theta_\text{phil}=0^\circ$ and $\theta_\text{phob}=180^\circ$\/.
Liquid accumulates at the upstream node until its volume is too
large to be held on the pattern by surface tension forces and a
drop detaches and flies off.
}
\end{figure}

\begin{figure}
\centerline{\includegraphics[width=9cm]{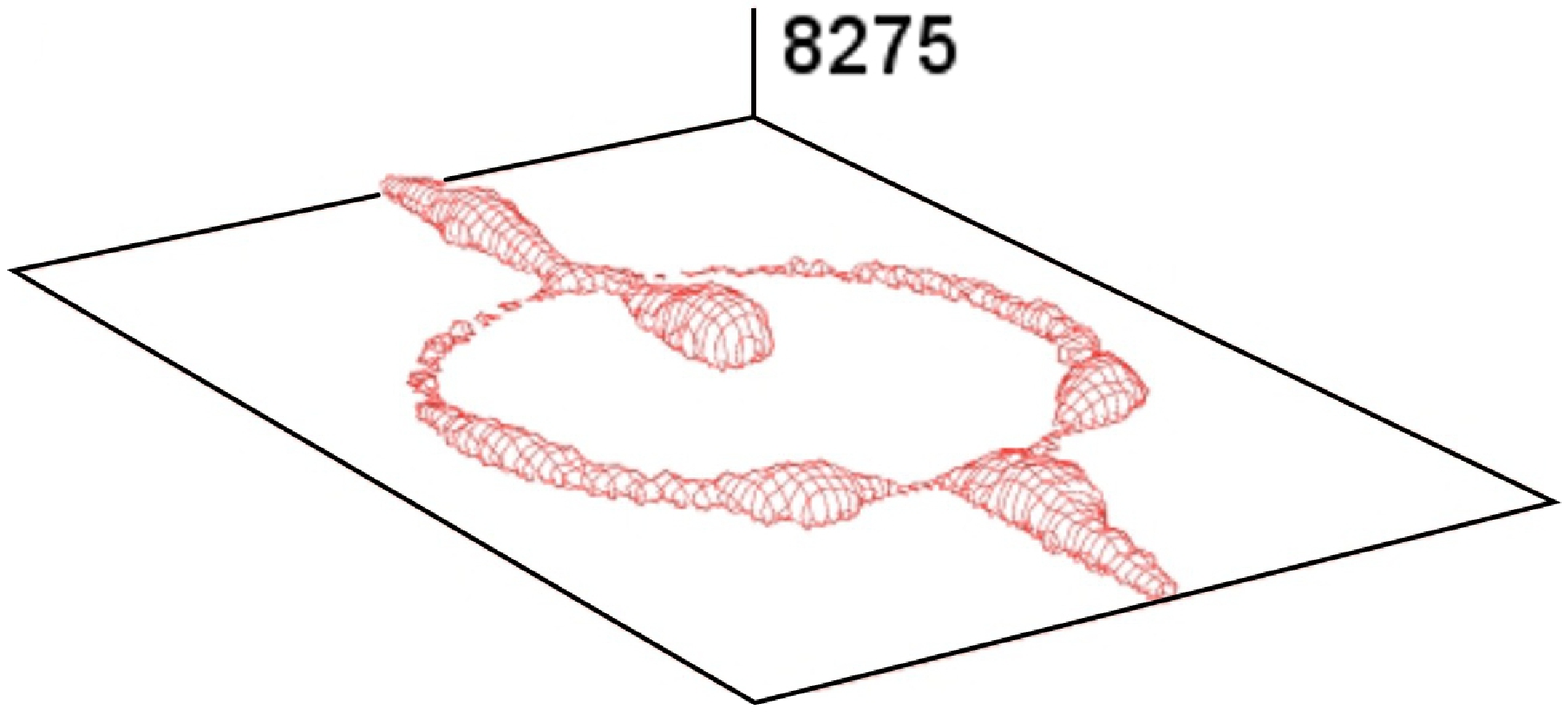}}
\caption{ \label{fig:md_big_pw}
Fluid behavior on the larger junction pattern of
Fig.~\ref{fig:md_big_nw} but with a {\em{partially}} wetting
exterior ($c_{liquid,rock}=0.75$);
$\theta_\text{phil}=0^\circ$ and $\theta_\text{phob}=90^\circ$\/.
The other simulation parameters are the same as in
Fig.~\ref{fig:md_open}.  In this case droplets of liquid detach
from the pattern and slide along the substrate.
}
\end{figure}

The number of fluid atoms in the open flow case shown in
Fig.~\ref{fig:md_open} is $119,840$ and the number of substrate
atoms is
$84480$\/. The box dimensions are identical to the those in the
shear flow case. With 269040 fluid and 345600 substrate atoms and 
box dimensions of $373.1\sigma \times 559.6\sigma \times
53.0\sigma$ the simulations leading to Figs.~\ref{fig:md_big_nw} and
\ref{fig:md_big_pw} are roughly 
twice as large.

The simulations are conducted in a box with periodic boundary
conditions 
in all three directions,
and the 
final
liquid temperature is fixed at 
$T=\epsilon/k_B$
by a Nos\'e-Hoover thermostat.
Initially, all atoms are placed on the sites of a $fcc$ lattice,
with the liquid on top of the wetting region having a rectangular
cross-section; the
temperature is ramped up to the 
above final value,
allowing the inner liquid to assume a circular cross-sectional
shape while remaining on top of the pattern.  If left alone, the
liquid forms pearls at the nodes of the junction pattern, due to
the same surface tension instability as observed in the LB case.
However, in the simulations discussed below a driving force is
applied before the pearls develop.  In the shear-cell case, we
translate the atoms in the top wall at constant velocity
$0.3\,\sigma/t_0$, while in the open case a body force of
magnitude $0.001\times m\,\sigma/t_0^2$ is applied to each atom
parallel to the long 
$x$-direction
of the substrate.

\begin{figure}
\centerline{\includegraphics[width=11cm]{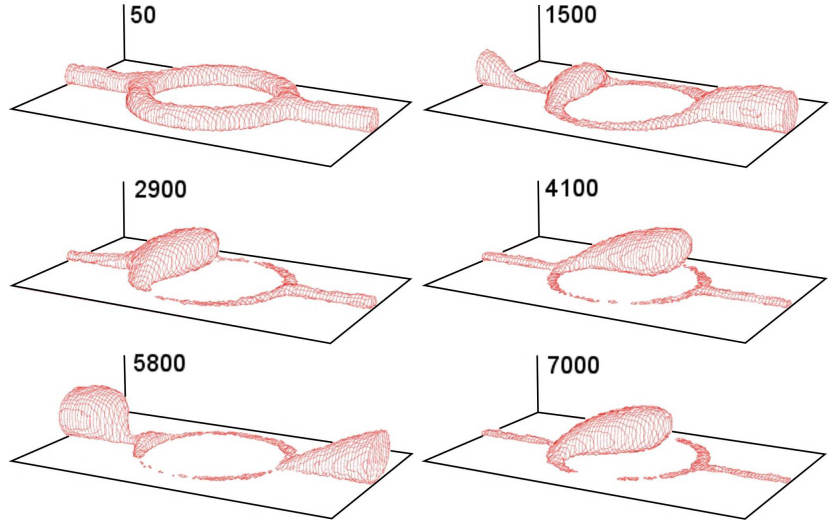}}
\caption{ \label{fig:md_closed} 
Sequence showing the evolution of a uniform liquid ridge (covered
by a second liquid) on top of a symmetrical junction pattern in a
shear cell (see Sect.~\ref{sec_md_methods} in the main text for a
full description of the initial configuration and the simulation
parameters;
$\theta_\text{phil}=0^\circ$ and $\theta_\text{phob}=180^\circ$)\/.
The spatial orientation of the sytem is the same as in
Fig.~\ref{fig_geo_lbe}(a), i.e., the chemical pattern lies in
the $xy$-plane and the 
sections are aligned with the $x$-direction.
The surface shown represents snapshots of the interface $\maA$
between the two liquids, and the motion is driven by translating
the top of the simulation box at a constant velocity
$0.3\,\sigma/t_0$ along the positive $x$-direction.
The frames are labeled by the time in units of $t_0$.
The straight lines indicate the boundaries of the simulation box.
}
\end{figure}

\subsection{Shear-cell flow}

The results of a simulation in rough correspondence to the LB
calculations are shown in Fig.~\ref{fig:md_closed}.  The figure
shows 
snapshots of
the interface between two immiscible
fluids in a shear cell, defined here as the surface on which the
number density of the inner fluid is $0.4\,\sigma^{-3}$, i.e.,
roughly half the 
bulk density of the inner fluid
at the same temperature.  The dimensions of the wetting pattern and
of the initial 
volume
of the inner fluid relative to the 
volume of the
simulation
box are in similar proportions as those in the symmetric LB case in
Fig.~\ref{fig_geo_lbe}(a)\/. Using the definitions for the
Reynolds 
number
and the Capillary 
number
in
Sec.~\ref{ssec_lbe_general}, we have Re $\approx 0.21$ and Ca
$\approx 0.56$ (compared to 
$2.5\lesssim 
\text{Re} 
\lesssim
3.5$ 
and 
$0.12\lesssim 
\text{Ca} 
\lesssim 0.16$
in the LB simulations, 
see also Sec.~\ref{sec_units} for a discussion)\/.
The corresponding values for the viscosity of the inner liquid and
for the liquid-liquid surface tension are 
$\eta^w\approx 2.12\,
m/(\sigma\,t_0)$ (obtained from a separate simulation of
Couette flow) and $\gamma\approx 2.50\,\epsilon/\sigma^2$
(found from a standard simulation of a slab of liquid
\cite{rowlinson02}),
respectively.
Furthermore, we emphasize that MD incorporates thermal fluctuations
while (our present version of) LB does not.

Nonetheless, the behavior of the two systems is rather similar.
The height of the inner liquid is initially uniform, and some
accumulation of water is visible at the nodes of the pattern before
the driving force is applied at $t=50\,t_0$.  The fluids move from
left to right (i.e., in the positive $x$-direction) in response to
the top boundary motion, and at $t=1500\,t_0$ one pearl of water
forms at the upstream node while a second one is traversing the
downstream periodic boundary.  Due to the symmetry of the wetting
pattern, the bulk of the upstream pearl is held in place as
additional water arrives through the periodic boundary, while small
amounts advance slightly along the circular arcs of the pattern.
As in the LB simulations, further symmetric advance along the arcs
is impossible due to the finite volume of liquid available, and
water continues to accumulate at the node until $t=2900\,t_0$,
after which a thermal fluctuation puts more material on the rear
arc and the pearl moves along it (see the snapshot at $4100\,t_0$).
Note the extreme overhang of the pearl -- only a narrow strip along
the arc is actually in contact with the solid.  Also, as in the LB
figures, the apparently water-free areas are a plotting artifact
due to those sampling bins having less than the water density
$0.4\,\sigma^{-3}$ which defines the interface as an 
isodensity surface.
At later times the single pearl reaches
and crosses the downstream node as well as the periodic boundary
($5800\,t_0$), halts temporarily at  the upstream node again, and
begins to follow the previous path along the rear arc
($7000\,t_0$).  In further repeated traversals of the pattern at
later times (not shown), the pearl configuration retains its shape
and (approximately) its volume, and always chooses the rear arc of
the pattern, because the thicker residue of water left behind there
after the first passage provides a less resistive path.


\subsection{Open flows}

Similar simulations have been performed on junction patterns with
only one 
one-component
liquid placed on top of the wetting
region, leaving the upper parts of the simulation box empty aside
from  a very dilute vapor resulting from evaporation of the liquid.
In Fig.~\ref{fig:md_open} we show  the motion 
of the interface
of a liquid-vapor system driven by a pressure gradient.
Explicitly, a body force is applied to each fluid atom parallel to
the pattern axis, as if the liquid was falling vertically under
gravity.  However, due to numerical limitations the magnitude of
the force had to be chosen to be much larger than terrestrial
gravity; a practical experimental procedure corresponding to our
simulations might use centrifugal forces on a rotating disk
with a radius of $1\,\text{cm}$ at $4\times
10^4~\text{rpm}$\cite{rauscher07a}.  One
qualitative change (note the frame at $100\,t_0$) is that the
interface is rougher with much more short length scale variation.
This may be understood intuitively as the result of molecules near
the surface having greater freedom of movement in the absence of a
viscous liquid ``cover,'' and more quantitatively as a
consequence of a lower surface tension for the 
liquid-vapor
interface in this simulation
($\gamma=0.49\,\epsilon/\sigma^2$)\cite{koplik06a}
in comparison to that of the 
liquid-liquid
interface
($\gamma= 2.50\,\epsilon/\sigma^2$)
considered above.
A second qualitative difference is that while in the previous
simulation there was always a single moving pearl, more diverse
morphologies of the interface
are seen here.
Initially, the fluid motion is similar to the shear-cell case:
liquid moves easily through the downstream node and 
the
periodic boundary while accumulating temporarily at the upstream node
(2050$\,t_0$), and then randomly choosing the rear arc to traverse
the junction (4725$\,t_0$).  However, on the second traversal, the
pearl instead chooses the front arc of the junction (7425$\,t_0$)
because the fluid film being on the front arc after the first
traversal through the 
rear
arc happens to be thicker than on
the 
rear
arc, perhaps again due to a random fluctuation.
At still later times, multiple pearls may be present and in
different states of motion.  For example at 11550$\,t_0$, there are
mobile pearls in both arcs and an accumulation at the upstream node
of the junction.  Subsequently, the bulge at the node merges
smoothly into the two pearls which continue to move, following each
other repeatedly through the rear arc (18750$\,t_0$).  Evidently,
the larger fluctuations present in the liquid-vapor case lead to a
less regular behavior.


The simulations shown so far have the attractive feature that the
water remains on top of the wetting pattern even in 
motion.  But
there are limitations to this behavior.  In
Fig.~\ref{fig:md_big_nw} we report on simulations of a larger
system in which the overall size of the simulation box and the
``skeleton'' of the wetting pattern (i.e., the length of the 
linear
sections and the radius of the centerline of the
arcs) are doubled in size, but the width of the wetting channel and
the initial height of the liquid are the same as before.  This
simulation thus involves twice the distance between nodes and twice
the volume of liquid.  As shown, in this case multiple pearls form
readily, which may be understood from a previous analysis of liquid
ridges on a linear wetting stripe \cite{koplik06a}, which revealed
a characteristic critical wavelength for the linear instability
leading to pearling behavior. Here, the longer intervals of 
linear
or curved channels exceed this wavelength.  A second
new feature is that if the liquid is held up at the upstream node
(due to symmetry) the additional liquid now present in the system
may build up to a pearl too large to be held in place by surface
tension forces until a fluctuation breaks the symmetry and an arc
of the pattern is selected at random.  In this situation, the pearl
is  simply pulled off the pattern by the driving forces and moves
downstream in the vapor.  In an alternative 
simulation, in which
the region outside the wetting pattern features partial
wetting ($\theta_\text{phob}\approx90^\circ$) rather than
nonwetting interactions with the liquid, as shown in
Fig.~\ref{fig:md_big_pw} streams of liquid leave the pattern and
move straight along the substrate in the direction of the force.
More generally, there is a competition between the wetting
interactions trying to hold the liquid on the pattern and the
applied forcing which attempts to move liquid downstream.  If the
strength of the forcing or the volume of liquid which accumulates
at a junction is too large, the driving force pushes the liquid
along the substrate irrespective of its wetting characteristics.


\section{Physical units}\label{sec_units}

Our simulation approach aims at reproducing
classical fluid dynamic behavior via (non-fluctuating) LB and 
at capturing nano-scale aspects via MD\/. These
two methods refer to different length scale regimes.  Those
regimes are specified by
the ratio of the Reynolds and the capillary number,
\begin{equation}
   \frac{\text{Re}}{\text{Ca}} = \frac{m\,\varrho\,\gamma}{(\eta^w)^2}\,w ,
\end{equation}
which, for given material parameters $m$, $\varrho$, $\gamma$, and
$\eta^w$, implies the 
value of the
channel width $w$ in
physical units.  Upon taking water under atmospheric pressure and
at room temperature as a reference system (i.e.,
$\gamma\approx70\,^{-3}\,\mathrm{kg}\,\mathrm{s}^{-2}$, $\eta^w\approx
10^{-3}\,\mathrm{kg}\,(\mathrm{m}\,\mathrm{s})^{-1}$, and
$m\varrho\approx 10^3\,\mathrm{kg}\,\mathrm{m}^{-3}$),
the ratios Re/Ca $\approx 21.7$ for LB and Re/Ca $\approx 0.38$ for
MD (shear-cell flow) imply $w\approx 310\,$nm for LB and $w\approx
5.4\,$nm for MD, and hence, the spatial resolution for LB
($w=6\,a$, see Fig.~\ref{fig_geo_lbe}(a))
and MD
($w=17.1\,\sigma$, see Sec.~\ref{sec_md_methods}) correspond to
\begin{equation}
a\approx51.7\,\mathrm{nm}\quad\text{and}\quad\sigma\approx0.32\,\mathrm{nm},
\end{equation}
respectively.
Likewise, the capillary numbers Ca $\approx 0.12$ for LB (low shear
rate) and Ca $\approx 0.56$ for MD imply shear velocities
$v_\text{shear}=\text{Ca}\,\gamma/\eta^w\approx
8.4\,\mathrm{m}/\mathrm{s}$
and
$v_\text{shear}\approx 39.2\,\mathrm{m}/\mathrm{s}$, respectively. 
The intrinsic time scale for our systems,
i.e., the typical time scale for thickness undulations of
wavelength $w$ on a thin film of this thickness\cite{willis10},
or of a rivulet on a channel of width $w$
is given 
by\cite{koplik06a}
\begin{equation}
   t^\star = \frac{\eta^w}{\gamma}w = \text{Ca}\,\frac{w}{v_\text{shear}}
   .
\label{t_star}
\end{equation}
Thus one has $t^\star\approx 4.4\,$ns for LB and $t^\star\approx 77\,$ps
for 
MD\/. The time resolutions of the LB and MD simulations are 
(see Eq.~\eqref{c_T_a}) $\Delta t=
a/(\sqrt{3}\,c_T)\approx440\,$ps and $t_0=
0.3\,\sigma/v_\text{shear} \approx2.4\,$ps, respectively.

\section{Free energy landscapes}\label{sec_freeEn}

\begin{figure}
\centerline{\includegraphics[width=11cm]{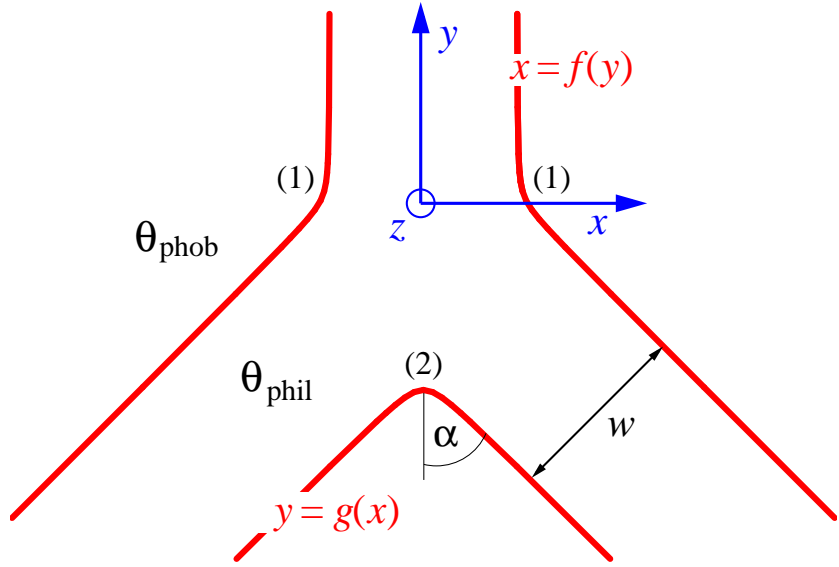}}
\caption{
The chemical pattern given by a Y-shaped (upside down) junction of
chemical channels with a uniform width $w$.  The channels are
defined by a wettability jump, i.e., a 
steplike
lateral variation
of the contact angle $\theta(x,y)$ between hydrophilic and
hydrophobic values $\theta_\text{phil}$ and $\theta_\text{phob}$.
The channel edges are given by the expressions
$x=f(y)=(b/2)[\sqrt{1+(y/b)^2}-y/b]\,\tan\alpha+w/2$
and $y=g(x)=-c\sqrt{1+(x/c^2}\,
\tan\alpha-w\,[1/\sin\alpha-1/(2\,\tan\alpha)]$
with the parameters $b=0.2$ and $c=0.2$ defining the curvature at
the corners (1) and (2), respectively.  The red lines indicate
$x=\pm f(y)$ and $y=g(x)$, the coordinate axes are drawn in blue,
and $\alpha$ is the opening angle of the junction (here
$\alpha = 45^\circ$), i.e., the pattern has a $yz$-mirror symmetry.
The origin lies in the mirror-symmetry plane at the potition
of maximal curvature of the channel edges, i.e., $f'''(y=0)=0$\/.
}
\label{fig_yjunc}
\end{figure}
\begin{figure}
\centerline{\includegraphics[width=11cm]{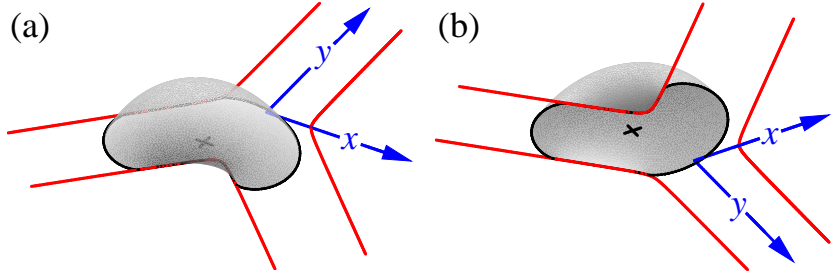}}
\caption{
Top view (a) and bottom view (b) of the interface configuration
$\maA$ of a non-volatile droplet in the vicinity of the
Y-shaped junction of chemical channels shown in
Fig.~\ref{fig_yjunc}, obtained numerically from Eq.~\eqref{mini}.
The three-phase contact line $\maL$ bounding $\maA$ is
drawn in black.  The droplet has a volume $V/w^3=0.75$, the
opening angle is $\alpha=45^\circ$, $\theta_\text{phil}=60^\circ$,
and $\theta_\text{phob}=180^\circ$.  The 
lateral
droplet position --
indicated by the black cross in (b) -- is specified via the
center of mass projection of $\maA$ onto the $xy$-plane,
$\bar{\bor}_\parallel=(\bar{x},\bar{y})=(-0.3w,-0.8w)$.
}
\label{fig_droplet_y}
\end{figure}

\begin{figure}
\centerline{\includegraphics[width=11cm]{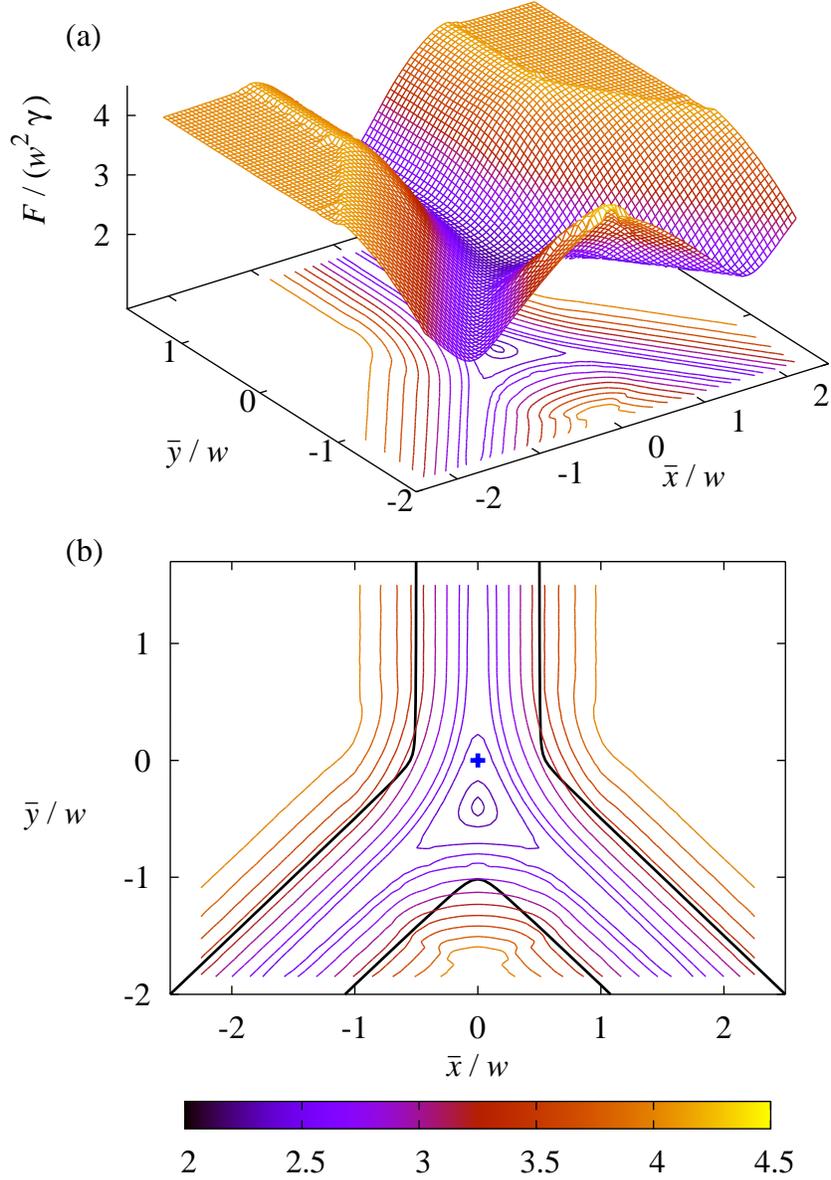}}
\caption{
The topography (a) and the corresponding contour plot (b) of
the free energy 
$F(\bar{x},\bar{y})$
as a function of the center of
mass coordinates $\bar{x}$ and $\bar{y}$ in the $xy$-plane for the
system shown in Fig.~\ref{fig_droplet_y}.  The blue cross within
the contour plot indicates the origin of the coordinate system (see
Fig.~\ref{fig_yjunc}) and the black lines correspond to 
the positions of the channel boundaries (see
Fig.~\ref{fig_yjunc})\/.  There is
a minimum of $F$ at $(\bar{x}=0,\bar{y}\approx-0.4\,w)$.
}
\label{fig_en}
\end{figure}
 
\begin{figure}
		\centerline{\includegraphics[width=11cm]{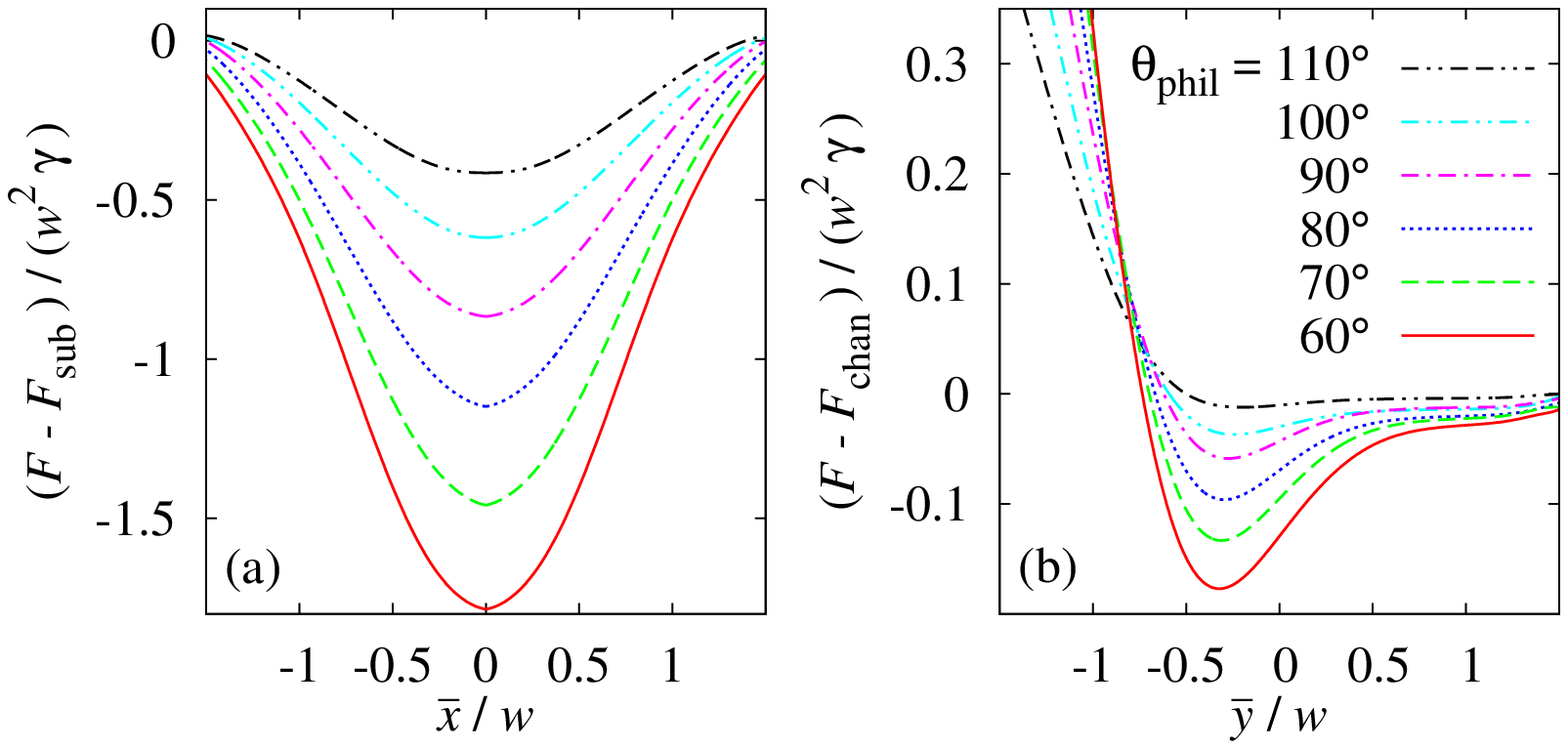}}
\caption{
Cuts of the free energy landscapes for $V/w^3=0.75$,
$\alpha=45^\circ$,
$\theta_\text{phob}=180^\circ$, and various hydrophilic contact
angles $\theta_\text{phil}\in[60^\circ,110^\circ]$, (a) along $\bar{x}$
through the global minimum and (b) along the
$\bar{y}\bar{z}$-symmetry plane at $\bar{x}=0$\/.  Due to the
symmetry of the system the global minimum always occurs at
$\bar{x}=0$\/.  In (a) and (b) the free energy is shown
relative to the
free energy of a droplet on a homogeneous hydrophobic substrate
$F_\text{sub}=F(\bar{x}\to\pm\infty,\bar{y})$ and
on a homogeneous straight chemical channel
$F_\text{chan}=F(|\bar{x}|<w,\bar{y}\to\infty)$, respectively.
Therefore
the values for the depths of the minima in (a) and (b) differ.
\label{fig_en_cut}}
\end{figure}

The previous analyses reveal that a key feature of motion through a
branched pattern is the behavior of droplets at a node.  In order
to gain further insight and to elucidate the dynamics associated
with this feature, we analyze a simplified quasi-static model which
focuses on the effects of surface tension (see
Figs.~\ref{fig_yjunc} and~\ref{fig_droplet_y}).  We consider a
non-volatile liquid "water" droplet near a node in contact both with
a substrate and a fluid "oil" or vapor phase and perform a
minimization of the interfacial free energy $\maF$ with respect to
the shape $\maA$ of the liquid-fluid interface configuration 
for steplike variations of the contact angle
under the constraints of a fixed volume and a fixed lateral
position of the center of mass of the droplet (see
Fig.~\ref{fig_droplet_y} and Appendix~\ref{app_F}).
For simplicity we assume the
liquid to be confined to the chemical channel, i.e.,
$\theta_{\text{phob}}=180^\circ$\/.

This requires to minimize the functional
\begin{equation}
   F = 
   \maF[\maA] 
	+ \Delta p\,C_p[\maA] 
	+ \bof_\parallel\cdot\boC_\parallel[\maA] .
\label{mini}
\end{equation}
The minimzation constraints are encoded in the scalar expression
$C_p$ and the vectorial expression $\boC_\parallel=(C_x,C_y)$ with
associated Lagrange multipliers $\Delta p$ and
$\bof_\parallel=(\text{f}_x,\text{f}_y)$, respectively.

The constrained minimization is carried out numerically using the
surface evolver package \cite{Brakke}\/. This is an adaptive finite
element algorithm, which evolves a certain triangulated, initial
interface configuration iteratively towards a state of minimal $F$
via a gradient projection method with $\Delta p$, $\text{f}_x$, and
$\text{f}_y$ as adjustable parameters.  

Since the droplet is confined to the chemical channel on
which the equilibrium contact angle $\theta$ is assumed to be
constant the interfacial free energy term in Eq.~\eqref{mini} has the form
\begin{equation}
   \maF[\maA] =
   \gamma\,
   \left(\int_{\maA}\!dA - \cos\theta
	\oint_{\maL}\!\!d\bol\cdot\hat{\bon}_y\,x\right) ,
\label{F}
\end{equation}
with the liquid-fluid interfacial tension $\gamma$ so that
$\gamma\,w^2$ provides the scale for the free energy $F$.  The
stripe width $w$ (see Fig.~\ref{fig_yjunc}) is the only lengthscale
within this model.  Young's contact angle $\theta$ measures the
wettability of the substrate, and $\hat{\bon}_y$ is the
$y$-component of the interface normal pointing outwards (see
Appendix \ref{app_F}).  The substrate surface is located in the
$xy$-plane and its tension with the liquid and the fluid
contributes via an oriented line integral along the boundary of
$\maA$, i.e., the three-phase contact line $\maL$
(Fig.~\ref{fig_droplet_y}). 
The chemical pattern is given by a steplike lateral
variation of the contact angle $\theta$ (see Figs.~\ref{fig_yjunc}
and~\ref{fig_droplet_y}).  

Non-volatility is ensured upon constraining the liquid volume to a
fixed value $V$\/.  This constraint is implemented via the Lagrange
multiplier $\Delta p$ so that
\begin{equation}
   \Delta p\,C_p = 
   \Delta p\,\left( V - \int_{\maA}\!d\boA\cdot
	\hat{\bon}_z\, z \right),
\label{vol_const}
\end{equation}
where $d\boA$ is the oriented surface element, and $\hat{\bon}_z$ is
the $z$-component of the outward normal of the interface.  The
Lagrange multiplier $\Delta p$ can be interpreted as the Laplace
pressure of the droplet.  

The constraint $\boC_\parallel$ is used in order to fix the center
of mass projection of the droplet onto the substrate plane at a
certain position $\bar{\bor}_\parallel=(\bar{x},\bar{y})$, i.e.,
\begin{align}
\label{bar_const}
   \bof_\parallel\cdot\boC_\parallel
   =\phantom{+}
   &\text{f}_x
   \left(\,
      \bar{x} - \frac{1}{2\,V}\int_{\maA} d\boA\cdot \hat{\bon}_x\,x^2\right)\\
   +
   &\text{f}_y
   \left(\,
      \bar{y} - \frac{1}{2\,V}\int_{\maA} d\boA\cdot \hat{\bon}_y\,y^2 \right)
    ,\nonumber
\end{align}
see Appendix \ref{app_com}.  The vectorial Lagrange multiplier
$\bof_\parallel=(\text{f}_x,\text{f}_y)$ can be interpreted as a
spatially constant, lateral force field, which suitably balances the
effect of surface tension forces.  In other words, the
configurational space $\{\maA\}$ for the minimization of the
free energy functional $\maF[\maA]$, which has been restricted to a subspace
of interface shapes with a certain volume $V$ by the contraint
$C_p$ given in Eq.~\eqref{vol_const}, is restricted further to the
subspace of shapes with a certain center of mass projection
$\bar{\bor}_\parallel$.  Naturally, this method only works for
droplets with a finite support, i.e., for
$\theta_\text{phil}>0^\circ$.

With this, the free energy $F$ of a droplet on the pattern can be
obtained as a function of $\bar{x}$ and $\bar{y}$ with $V$ and
the wettability contrast $\theta_\text{phob}-\theta_\text{phil}$ as
input parameters (see Fig.~\ref{fig_en} for the topography and a
corresponding contour plot of $F(\bar{x},\bar{y})$)\/.  Naturally,
the chemical channels translate into energetic valleys with a
global minimum at $(\bar{x}=0,\bar{y}\approx-0.4\,w)$ near the
intersection point of the channel midlines at
$\bar{y}=\frac{w}{2}\left(1-(\sin45^\circ)^{-1}\right)\approx-0.2\,w$.
Upon decreasing the wettability contrast the topography of the free
energy landscape becomes less structured: the depth of the
global minimum relative to the free energy of a droplet far from the
chemical channels as well as relative to the free energy of a
droplet on a straight channel decreases (see
Figs.~\ref{fig_en_cut}(a) and (b), respectively). 
However, the position of the global minimum
does not shift (see Fig.~\ref{fig_en_cut} (b)). The
$\bar{y}$-position of the minimum certainly depends on the opening
angle $\alpha$.  For $\alpha=60^\circ$, i.e., for a threefold
rotational symmetry of the junction, it is
right at the intersection point of the channel midlines.

The occurrence of the global free energy minimum at the pattern
junction implies a certain qualitative behavior of the droplet: An
unconstrained droplet -- given by $\bof_\parallel\equiv0$ within
Eq.~\eqref{mini} -- would be sucked in and trapped at the junction.
Equivalently, a trapped droplet has to overcome a free energy
barrier in order to escape from the junction.  In this resect it is
worthwhile to compare this with the LB-results in
Figs.~\ref{fig_a80_v02}-\ref{fig_a30}, which show preferential
droplet formation on the junction during pearling, as well as the
hang-up and merger of driven droplets there. In particular
Fig.~\ref{fig_a30}(a) shows a permanent hang-up for
$\theta_\text{phil}\approx 30^\circ$.
 
The contour plot in Fig.~\ref{fig_en}(b) shows, that the
consequences of the free energy minimum, i.e., leading to a
possible droplet hang-up, can be avoided by guiding the droplet via
an additional external, lateral force in such a way that its center
of mass projection $\bar{\bor}_\parallel$ moves along an open
contour line.

\section{Summary, conclusion, and outlook} \label{sec_concl}

We have analyzed theoretically the forced motion of liquid rivulets
through branched chemical patterns on a substrate
(Figs.~\ref{fig_geo_lbe} and~\ref{fig_cuts_lbe}).  The purpose of
this study is to demonstrate the feasibility of this set-up for
controlled fluid transport, which might serve as a component of
future ``lab-on-a-chip'' devices.  Our calculations employ a
lattice Boltzmann method (Fig.~\ref{fig_d3q19}) for the isothermal
description of microscale systems and molecular dynamics for their
nanoscale counterparts.

The microscale simulations (see Figs.~\ref{fig_a80_v02} -
\ref{fig_a30}) show that within reasonable limitations for the flow
rate and the liquid volume, the wetting pattern is capable to
direct liquid motion along desired paths.  Familiar surface tension
instabilities lead to pearl formation at the nodes of the pattern
(see in particular Fig.~\ref{fig_a80_v02}), but as in simpler flow
configurations \cite{koplik06a,rauscher07a} the pearls are mobile
and move along the pattern, and therefore need not to obstruct the
motion of the liquid.  Although the nanoscale systems are
particularly sensitive to thermal fluctuations, the behavior of the
liquid in this domain is nonetheless fairly similar (see
Figs.~\ref{fig:md_closed}-\ref{fig:md_big_pw}).

Concerning the possibility to control these systems the challenge
is to direct the liquid preferentially along one of multiple paths
leaving a node.  Our microscale simulations indicate that small
variations in the width of the wetting channels can suffice for
this purpose (see Figs.~\ref{fig_a80_v03}(b)
and~\ref{fig_a30}(b)).  The liquid prefers to move along a wider
channel where viscous resistance is lower.  Other mechanisms for
directing flow, such as imposed electric fields acting on an ionic
liquid, bursts of air directed towards one side of a node, smooth
wettability gradients on the surface, etc., might work as well or
better.  We leave these possibilities for future studies.

For a qualitative understanding of the behavior of driven pearls at
the nodes we have employed a quasi-static model based on the action
of surface tensions.  This yields the free energy landscape with a
state of minimal free energy for an isolated droplet sitting at a
node (see Figs.~\ref{fig_yjunc} - \ref{fig_en_cut}).

Likewise, an additional consideration for future work is the
behavior of suspensions or complex liquids, e.g., lyotropic or
thermotropic liquid crystals, flowing along chemical patterns.

\appendix

\section{Free energy}\label{app_F}

Using the local version of Young's law,
$\gamma\,\cos\theta(x,y)=\gamma_{rf}(x,y)-\gamma_{rw}(x,y)$ with the 
indices $r$, $w$, and $f$ for "rock", "water", and "fluid" (i.e.,
"oil" or vapor), respectively, within
the capillary model for a chemically heterogeneous substrate domain with a
laterally varying macroscopic contact angle $\theta(x,y)$ 
the free energy $\maF$ can be expressed as
\begin{equation}
   \frac{\maF}{\gamma} = 
   \int_{\maA}\!dA
   -
   \int_{\maS}\!d\boA\cdot\hat{\bon}_z \,\cos\theta(x,y)
\end{equation}
with 
the liquid-fluid interfacial tension $\gamma$
as a scaling paramter. Here we assume that the substrate surface
and thus also the "water"-substrate interface $\maS$
lie in the
$xy$-plane and that $\maS$ is oriented such that its normal
vector is $\hat{\bon}_z=(0,0,1)$\/. 

The integral over $\maS$ can be converted into an
integral over the three-phase contact line 
if one finds a vector potential for the vector 
field $\hat{\bon}_z\,\cos\theta(x,y)$\/. A 
simple (and for the present geometry convenient) choice of
the corresponding gauge is adopted by
\begin{equation}
   \hat{\bon}_z \,\cos\theta(x,y) = 
	\bm{\nabla}\times\left[\hat{\bon}_y\,C(x,y)\right],
\end{equation}
with $\hat{\bon}_y=(0,1,0)$  and 
$C(x,y)=\int^x_{x_0} dx'\,\cos\theta(x',y)$\/. (Note that
$\bm{\nabla}\times [\hat{\bon}_y \,C(x,y)]$ is independent of
$x_0$\/.) Applying Stokes' theorem yields
\begin{equation}
   \frac{\maF[\maA]}{\gamma} = 
     \int_{\maA}\!dA - 
     \oint_{\maL}\!d\bol\cdot\hat{\bon}_y\,C(x,y),
	  \label{heterogeneous}
\end{equation}
where the three-phase contact line $\maL$ is the oriented boundary
of $\maS$ as well as of $\maA$ (see Fig.~\ref{fig_droplet_y}).

If a chemical pattern with a steplike variation of the local
equilibrium contact angle $\theta(x,y)$ confines the drop to a
chemically homogeneous part of the substrate 
Eq.~\eqref{heterogeneous} reduces to Eq.~\eqref{F}\/. This holds
even if a portion of $\maL$ coincides with an interval of the
boundary of the chemical channel.

\section{Center of mass constraint}\label{app_com}

In order to constrain the lateral position of the center of
mass $V^{-1}\,\int_{V}dV\,\bor$ of a droplet 
of volume $V$ and of homoegeneous density
to $(\bar{x},\bar{y})$ one adds the expression
\begin{align}
   \bof_\parallel\cdot\boC_\parallel
   = \phantom{+}
   &\text{f}_x\left( \bar{x} - \frac{1}{V}\int_{V} dV\,x \right)\nonumber\\
   +&\text{f}_y\left( \bar{y} - \frac{1}{V}\int_{V} dV\,y
	\right)\label{cmosconst}
\end{align}
with the Lagrange multiplier $\bof_{||} = (f_x,f_y)$ to the free
energy functional. The volume integrals in Eq.~\eqref{cmosconst}
can be converted into surface integrals using Gau{\ss}' theorem. To
this end one expresses each of the scalar integrands as the
divergence of a suitable vector field. For droplets
residing on a planar substrate which lies in the $xy$-plane a
convenient choice is
\begin{equation}
x = \bm{\nabla}\cdot (\tfrac{1}{2} \,
x^2\,\hat{\bon}_x)\quad\text{and}\quad
y = \bm{\nabla}\cdot (\tfrac{1}{2}\,y^2\, \hat{\bon}_y).
\end{equation}
Since the area element $d\boS=dS\,\hat{\bon}_z$ of $\maS$ is
orthogonal to $\hat{\bon}_x$ and $\hat{\bon}_y$ the contribution of
the surface integral over the liquid-substrate interface
$\maS$ to the surface integral over the total surface of $V$ 
vanishes and the term $\bof_\parallel\cdot\boC_\parallel$
in Eq.~\eqref{mini} reduces to the expression given in
Eq.~\eqref{bar_const}\/.

\section*{Acknowledgements}
Computational resources were provided by the Rechenzentrum Garching
der Max-Planck-Gesellschaft und des Instituts f{\"u}r Plasmaphysik and
the Scientific Supercomputing Centre Karlsruhe.


\providecommand*{\mcitethebibliography}{\thebibliography}
\csname @ifundefined\endcsname{endmcitethebibliography}
{\let\endmcitethebibliography\endthebibliography}{}

\end{document}